# Universal Path Gain Laws for Common Wireless Communication Environments

Dmitry Chizhik, *Fellow, IEEE*, Jinfeng Du, *Member, IEEE*, and Reinaldo A. Valenzuela, *Fellow, IEEE*

*Abstract*— **Simple and accurate expressions for path gain are derived from electromagnetic fundamentals in a wide variety of common environments, including Line-of-Sight (LOS) and Non-Line-of-Sight (NLOS) indoor, urban canyons, urban/rural macro, outdoor-indoor and suburban streets with vegetation. Penetration into a scattering region, sometimes aided by guiding, is the "universal" phenomenon shared by the diverse morphologies. Root Mean Square (RMS) errors against extensive measurements are under 5 dB, better than 3GPP models by 1-12 dB RMS, depending on environment. In urban canyons the models have 4.7 dB RMS error, as compared to 7.9 dB from linear fit to data and 13.9/17.2 dB from LOS/NLOS 3GPP models. The theoretical path gains depend on distance as a power law with exponents from a small set {1.5, 2, 2.5, 4}, specific to each morphology. This provides a theoretical justification for widely used power law empirical models. Only coarse environmental data is needed as parameters: street width, building height, vegetation depth, wall material and antenna heights.**

## I. Introduction

Performance of wireless communications is critically dependent on achieving adequate coverage, particularly challenging at mm wave bands. This paper concentrates on modeling path gain, the most basic aspect of propagation, key in determining coverage. Path gain is defined as the ratio of average receive and transmit powers for omnidirectional, co-polarized transmit/receive antennas. Models are needed both to estimate system performance and requirements in generic circumstances, such as determining inter-site distance required for adequate coverage at a given frequency and transmit power as well as planning and placement of base stations in a particular area.

Prominent approaches to propagation modeling in communications include ray-tracing [1][2] as well as empirical models [3]-[15] derived from field measurements for various deployment scenarios such as indoor office and urban street canyons. Ray tracing can handle very general environments to predict the full impulse response of the channel but requires detailed environmental information, such as individual building walls. In outdoor-indoor scenarios, this includes the need to specify interior walls, which play a key role in indoor scattering. Furniture information is generally unavailable and is usually ignored despite being many wavelengths across. Vegetation and urban street clutter play a critical role in attenuating signal, especially at mm wave bands, with attenuation through a 10 m crown of a single tree on the order of 20 dB at 28 GHz. Such detailed information is often difficult to obtain.

Widely used empirical models, such as 3GPP [3] and WINNER [4], specify path gain as a function of range for several important scenarios, such as urban microcell, urban canyons, rural and indoor. Path gain $P$ is expressed (in dB) as a function of range $r$ (m) in the "slope-intercept" form in terms of distance exponent $n$ and 1-m intercept $P_1$:

$$P_{\mathrm{dB}} = 10\log_{10}\left(\frac{P_1}{r^n}\right) = 10\log_{10} P_1 - 10n\log_{10} r \qquad (1)$$

No environmental details are required, only generic morphological label, like "urban microcell". These models are used as references in current work. Known theoretical examples of form (1) include Friis free space formula with $n$=2 and LOS propagation between antennas of heights $h_1$ and $h_2$ above a ground plane of reflection coefficient -1, with $n$=4 beyond the breakpoint $r = 4h_1h_2/\lambda$ for wavelength $\lambda$.

Deployment of cellular communication systems is often preceded by an extensive path gain measurement campaign, from which $P_1$ and $n$ specific to the area are determined from Least Means Square (LMS) fit to the data. The RMS goodness of fit is regarded as shadow fading.

Extensive measurements in both suburban [16] and urban [17] environments have shown wide variation in path gain among subsets of data nominally belonging to the same morphological class, e.g. urban canyon measurements [17] have shown street-street variation of path loss at similar distance spanning a range of 30 dB. Since the empirical path loss models are generally functions of distance and LOS/NLOS status alone, they do not account for such variations. It is of interest to develop models adaptable to local conditions to improve prediction accuracy for coverage planning.

Most desirable models are a) accurate, b) dependent only on reasonably available environmental parameters and c) simple to implement. Closed form expressions (as in empirical models [3][4]) are attractive in that they are easy to communicate and do not require complex programming to be developed and supported. At the same time, reducing or eliminating site-specific measurements is highly desirable. The Walfisch-Bertoni model [18] is an example of an analytical model with a relatively simple path gain expression, obtained from a numerical solution of over-rooftop propagation.

In this work we derive simple yet accurate path gain expressions from electromagnetic fundamentals for important canonical environments, including urban canyons with and

Dmitry Chizhik, Jinfeng Du and Reinaldo A. Valenzuela are with Nokia Bell Laboratories, Murray Hill, NJ 07974, USA (e-mail: dmitry.chizhik@nokia-bell-labs.com, jinfeng.du@nokia-bell-labs.com, reinaldo.valenzuela@nokia-bell-labs.com).

without trees, outdoor-indoor in canyons, indoor corridors (LOS and NLOS), traditional urban/rural macro cellular propagation from above roof-top base to below rooftop/treetop terminal and suburban propagation through foliage. Notable aspects are:

- Under 5 dB RMS error against measurements;
- Derived path gain depends on distance as a power law, justifying the widely used slope-intercept form (1) in empirical models;
- Morphology dependent distance exponents $n$ of 1.5 for LOS in canyons, 2.5 for canyon outdoor-indoor (or street-cluttered sidewalk), 4 for in unguided penetration into a cluttered space (indoors, behind foliage, etc.);
- 1-m intercept $P_1$ in (1) is found dependent on general environmental parameters, such as street width, antenna height and placement in the street, vegetation depth and wall material properties.

Derivations of the approximate reflection coefficient from rough surface and propagation from free space into diffuse half-space, which are an important part of this work, are put into appendices to make the paper more readable.

Average path gain between omnidirectional antennas, without multipath fading, is the main channel property modeled and compared against data. The impact of multipath fading on narrowband (CW) measurements is reduced through averaging, either locally over several wavelength (at 2 GHz, 3.5 GHz) or over angle when using a spinning horn antenna (at 28 GHz). Once the multipath fading is removed, the resulting average power corresponds to an average over frequency as well. 28 GHz measurements collected with a spinning directional horn, were averaged over all azimuth directions, resulting in equivalent azimuthally omnidirectional antenna [22].

## II. General problem formulation

We are interested in modeling fields in canonical environments arising in radio communications. Here the quantity of interest is the received power when both transmit and receive antennas are electrically small (i.e. “omnidirectional”). Fields from such antennas for both polarizations are related to corresponding Hertz potentials [19], each proportional to the scalar Green’s function $G(\mathbf{r},\mathbf{r}_s)$, which is the field at $\mathbf{r}$ due to a point source at $\mathbf{r}_s$. In the time-harmonic case, it satisfies the scalar Helmholtz equation:

$$\nabla^2 G(\mathbf{r},\mathbf{r}_s) + k^2 G(\mathbf{r},\mathbf{r}_s) = \delta(\mathbf{r}-\mathbf{r}_s) \tag{2}$$

where wavenumber $k = 2\pi/\lambda$ and the point source is represented by the Dirac delta function $\delta(\mathbf{r}-\mathbf{r}_s)$. Directional antenna response may be constructed from elementary solutions to (2). Cross-polarization coupling, expected due to scattering, is not addressed here, noting that measured values of median cross-polarization power ratio have been reported as below -8 dB in environments of interest [3]. Inhomogeneous media are typically represented by adding boundary conditions to (2), introducing reflections and scattering, as is done in subsequent sections, for several environments. In free space, the solution to (2) is the spherical wave:

$$G_0(\mathbf{r}_s,\mathbf{r}) = \frac{e^{ik|\mathbf{r}_s-\mathbf{r}|}}{4\pi|\mathbf{r}_s-\mathbf{r}|} \tag{3}$$

The corresponding power received when both transmit and receive antennas are unit gain is given by:

$$P_0 = \lambda^2 \left|G_0(\mathbf{r}_s,\mathbf{r})\right|^2 = \frac{\lambda^2}{(4\pi)^2\left|\mathbf{r}_s-\mathbf{r}\right|^2} \tag{4}$$

recognized as the Friis equation [19].

## III. LOS Propagation in a Street Canyon or Indoor Corridor

This section considers propagation with transmit and receive antennas placed in a street canyon or indoor corridor, bounded by walls on either side and the ground. Propagation in such an environment is modeled here as a combination of the direct path as well as reflections from the walls and the ground. Ceiling reflection, relevant for an indoor corridor, is neglected as under-ceiling ductwork and cabling, common in commercial buildings, are presumed to scatter incident signals into steep elevation and azimuth angles, to be scattered many times and eventually absorbed in lossy materials, like the concrete floor.

The source and receiver are both taken to be in the middle of the canyon. Extension to off-center antennas is discussed at the end of the section. Using the image theory, various order reflections may be represented as emanating from a sequence of images, shown in Fig 1. Here the $x$-axis is along the canyon, $y$-axis across the canyon and $z$ is the vertical coordinate. For a source at $\mathbf{r}_s = (0,0,z_s)$ in the middle of the waveguide of width $w$, the $m^{th}$ order reflection appears as emanating from an image source a distance $mw$ from the center of the canyon, with coordinates $\mathbf{r}_m = (0, mw, z_s)$.

Neglecting the ground reflection for the moment, the scalar Green’s function (point source response) at a receiver at $\mathbf{r} = (x,0,z)$ for a sum of images corresponding to reflections from the canyon walls:

$$G(\mathbf{r}_s,\mathbf{r}) = G_0(\mathbf{r}_s,\mathbf{r}) + \sum_{m=1}^{\infty} G_m(\mathbf{r}_m,\mathbf{r}) + \sum_{-\infty}^{m=-1} G_m(\mathbf{r}_m,\mathbf{r}) \tag{5}$$

where the free space Green’s function $G_0(\mathbf{r}_s,\mathbf{r})$ is given by (3) and the $m^{th}$ order reflection response due to image at $\mathbf{r}_m$ is

$$G_m(\mathbf{r}_m,\mathbf{r}) = \Gamma_m^{|m|} \frac{e^{ik|\mathbf{r}_m-\mathbf{r}|}}{4\pi|\mathbf{r}_m-\mathbf{r}|} \tag{6}$$

The sums over negative and positive $m$ account for images on the either side of the canyon in Fig. 1. The wall reflection coefficient $\Gamma_m$, raised to the $|m|^{th}$ power to account for $m$ reflections, is dependent on the wall grazing angle $\theta_m$

$$\theta_m = \tan^{-1}(mw/r) \approx mw/r \tag{7}$$

with the distance between transmitter and receiver

$$r \equiv \sqrt{x^2 + \left(z_s - z\right)^2} \tag{8}$$

The average received power

$$P = \lambda^2 \left\langle \left| G(\mathbf{r}_s, \mathbf{r}) \right|^2 \right\rangle \tag{9}$$

often obtained through local spatial average, is the sum of powers of the terms in (5)

$$P \approx \lambda^2 \sum_{m=-\infty}^{\infty} \frac{\left|\Gamma_m\right|^{2|m|}}{(4\pi)^2 \left|\mathbf{r}_m - \mathbf{r}\right|^2} \tag{10}$$

with $m = 0$ corresponding to direct path.

For $r \gg |m| w$, $|\mathbf{r}_m - \mathbf{r}|$ is expanded in terms of $r$ in (8):

$$\left|\mathbf{r}_m - \mathbf{r}\right| = \sqrt{x^2 + (mw)^2 + \left(z_s - z\right)^2} \approx r + \frac{m^2 w^2}{2r} \tag{11}$$

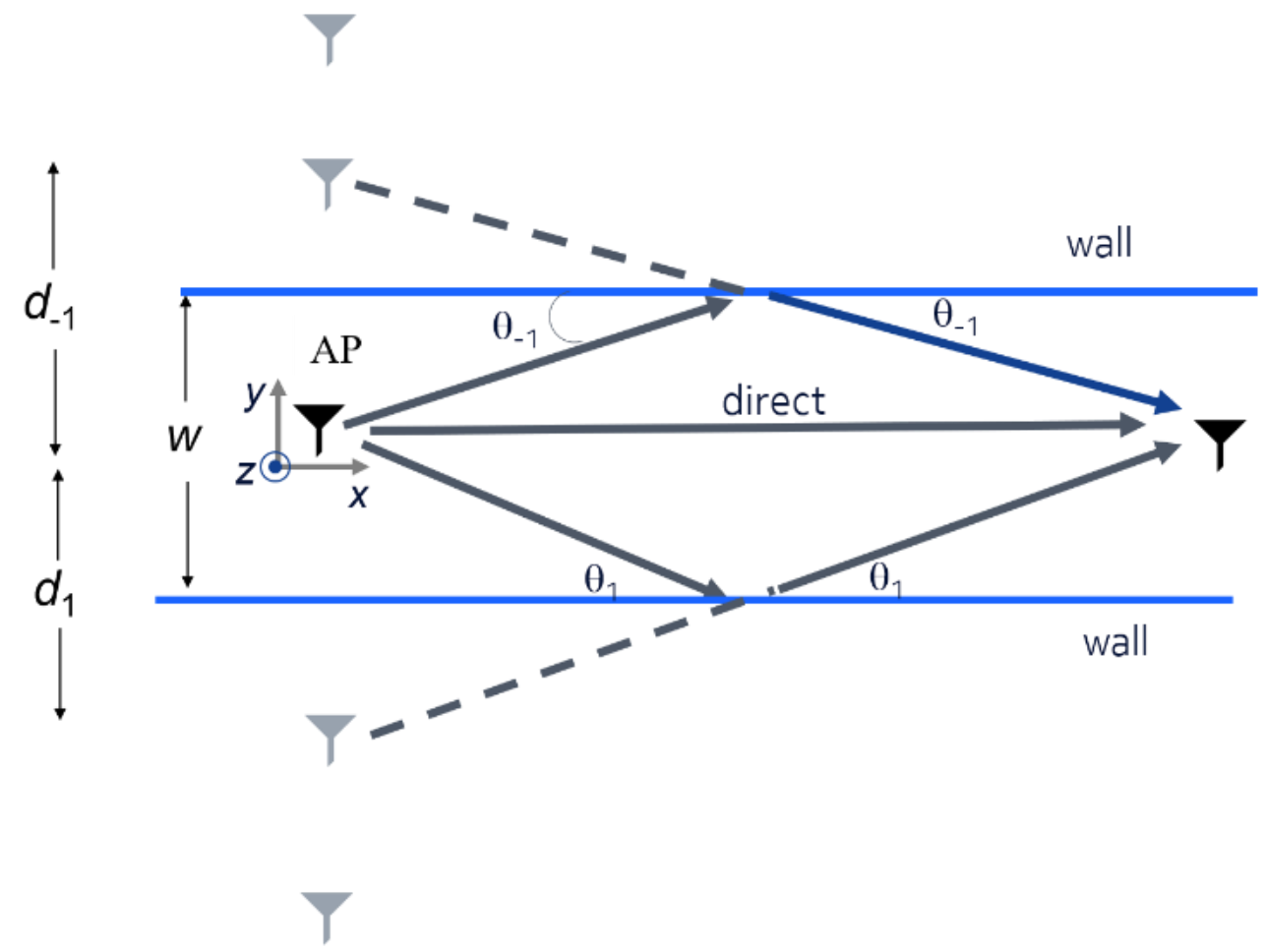

Fig. 1. Top view of canyon of width $w$, with an infinite sequence of source images (in grey) representing wall reflections.

Similarly, using Taylor series expansions to 2nd order,

$$\frac{1}{\left|\mathbf{r}_m - \mathbf{r}\right|} = \frac{1}{\sqrt{r^2 + (mw)^2}} \approx \frac{1 - (mw)^2 / 2r^2}{r} \approx \frac{e^{-(mw)^2/2r^2}}{r} \tag{12}$$

Terms with $|m|$ large enough to violate the assumption $r \gg |m| w$ are attenuated by reflection loss $\left|\Gamma_m\right|^{2|m|}$ in (10).

Using (12) in (10), leads to

$$P \approx \left(\frac{\lambda}{4\pi r}\right)^2 \sum_{m=-\infty}^{\infty} \left|\Gamma_m\right|^{2|m|} e^{-(mw)^2/r^2} \tag{13}$$

Contribution of the $m^{\text{th}}$ order reflection to total average power is attenuated by the factor $\left|\Gamma_m\right|^{2|m|}$ due to $m$ wall reflections and by the excess spreading loss factor $e^{-(mw)^2/2r^2}$ corresponding to path length that increases with $m$.

Using the approximation (51) in Appendix I to the reflection coefficient for incidence on a wall at small grazing angle (7)

$$\left|\Gamma_m\right|^{2|m|} \approx \left|e^{-L|\theta_m|/2}\right|^{2|m|} = e^{-L|m\theta_m|} = e^{-Lm^2 w/r} \tag{14}$$

in terms of the unitless wall-loss parameter $L$

$$L \equiv 4/n_{\text{eff}} + 32k^{3/2} A^2 p_1 p_2 \sqrt{\mu_1 + \mu_2} \tag{15}$$

The two terms in (15) represent reflection loss from a smooth dielectric wall with refraction index $n_{\text{eff}}$, and loss due to scatter from wall roughness, respectively. “Roughness” considered (appendix I) is due to window/door wells in a wall, with window well half-depth $A$, relative wall fraction occupied by the window/door $p_1$, rest of the wall $p_2 = 1 - p_1$, and average window/door width $1/\mu_1$ and inter-window spacing $1/\mu_2$.

Substituting (14) into (13), the sum over reflection orders $m$ in (13) is viewed as a rectangular-rule approximation for an integral over a continuous variable $s$:

$$\sum_{m=-\infty}^{\infty} e^{-m^2(L+w/r)w/r} \approx \int_{-\infty}^{\infty} ds e^{-s^2(L+w/r)w/r} = \sqrt{\frac{\pi r}{(L+w/r)w}} \approx \sqrt{\frac{\pi r}{Lw}} \tag{16}$$

where the last approximation is justified for $L \gg w/r$, i.e. excess spreading loss is negligible compared to reflection loss, e.g. for non-metallic walls and $r \gg w$. Using (14) and (16), path gain (13) in LOS canyon becomes

$$P_{\text{LOS canyon}} \approx \frac{\lambda^2}{16\pi^{1.5}\sqrt{wL}\; r^{1.5}} \tag{17}$$

The distance $r$ is raised to the exponent of 1.5, independent of any environmental variable. Exponent of 1.5 was found [32] in ocean acoustic waveguides using a different approach. In (17) the dependence of the 1-m “intercept” of path gain $10\log_{10}\left(\lambda^2 / 16\pi^{1.5}\sqrt{wL}\right)$ (dB) on wall materials through $n_{\text{eff}}$ and canyon width $w$ is relatively weak, as changing either by a factor of 2 changes the intercept by less than 1.5 dB.

Ground (floor) reflection with field reflection coefficient $\Gamma_g$ is now included by adding an image of the source to (17):

$$P \approx \frac{\lambda^2}{16\pi^{1.5}\sqrt{wL}\; r^{1.5}} \left| e^{ikr} + \Gamma_g e^{ikr_g} \right|^2 \tag{18}$$

where $r_g$ is the distance from the source image in the ground to the receiver. As done for wall reflections in Appendix I, the ground reflection coefficient dependence on the ground grazing angle $\theta_g = \sin^{-1}\left(z_s + z\right)/r_g$ may be approximated for vertical polarization by $\Gamma_g \approx -e^{-\left(2n_g^2/\sqrt{n_g^2-1}\right)\theta_g}$ (45) for low grazing angles, with typical ground refraction index (concrete/dry soil) $n_g \approx \sqrt{5} \approx 2.2$. At ranges before the breakpoint $r < 4z_s z/\lambda$, (18) produces a two-ray beating pattern with varying range $r$. Average power in that regime is obtained by summing the powers of direct and reflected arrivals, leading to

$$P_{\text{LOS corridor}} \approx \frac{\lambda^2}{16\pi^{1.5}\sqrt{wL}\; r^{1.5}}\left(1+\left|\Gamma_g\right|^2\right) \quad (19)$$

As range $r$ increases, $\left|\Gamma_g\right|^2 \to 1$.

Predictions (19) are compared against LOS corridor measurements of path gain at 2 GHz in Fig. 2. The data was collected using omni antennas in an office corridor $w$=1.6 m wide, with transmitter 2.2 m above the floor and receiver at 1 m. Measurement details are in [21]. Parameters defining wall loss $L$ (15) are $n_{\text{eff}} = 1.7$, $A$=0.035 m, $p_1 = 0.25$, $p_2 = 0.75$, $1/\mu_1 = 1$ m, $1/\mu_2 = 3$ m. Path gain (19) has 2.8 dB RMS deviation from data, slightly greater than 2.6 dB obtained from linear fit to data. Including ground reflection coherently as in (18) apparently reproduces the observed beating pattern in Fig. 2 enough to reduce the RMS deviation to 2.4 dB.

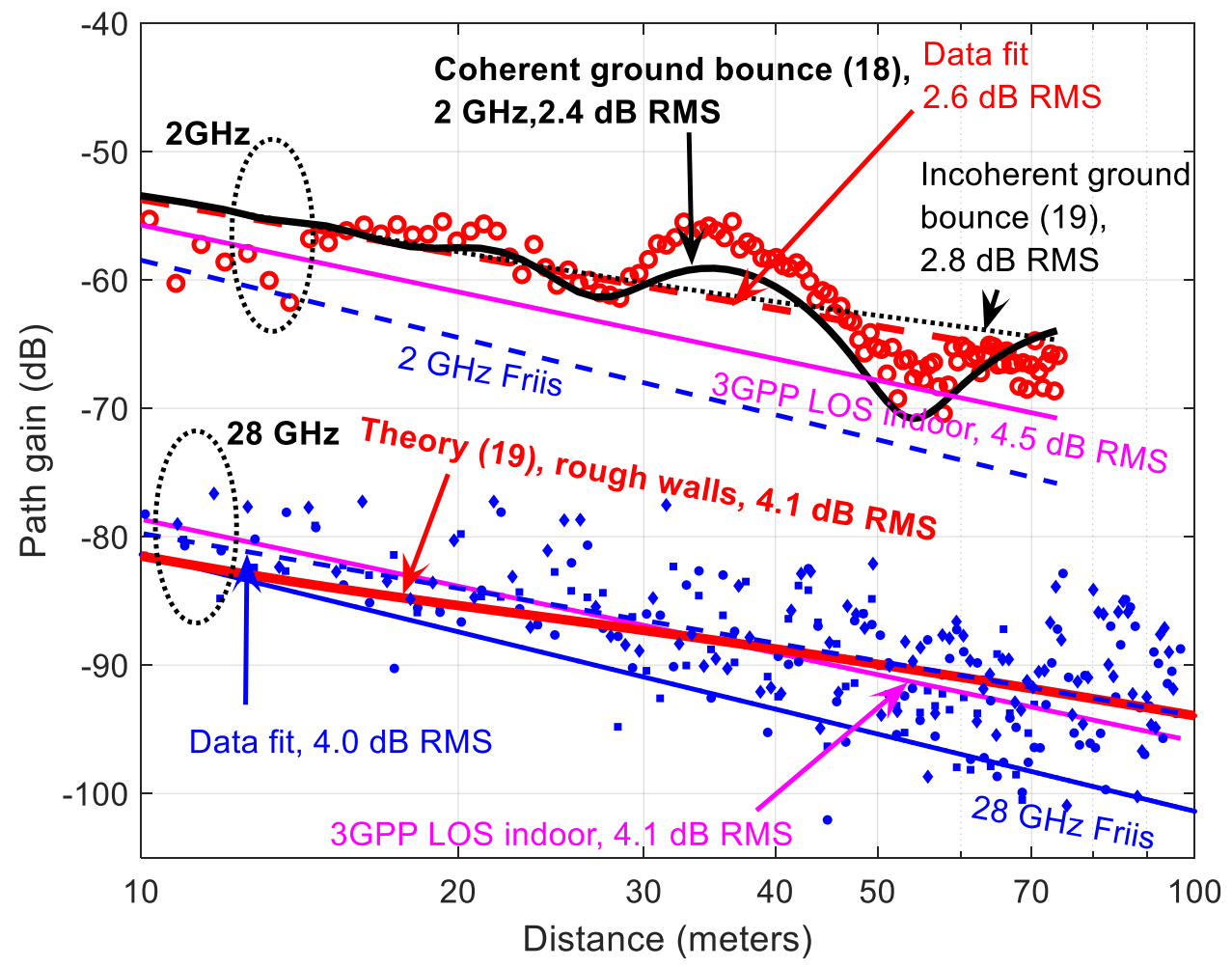

Fig. 2. Path gain measured and modeled at 2 GHz and 28 GHz in a LOS office corridor. "Coherent" and "incoherent" ground bounce theory are (18) and (19), respectively. Wall loss $L$ (15) parameters: $n_{\text{eff}} = 1.7$, $A$=0.035m, $p_1 = 0.25$, $p_2 = 0.75$, $1/\mu_1 = 1$ m, $1/\mu_2 = 3$ m.

Single tone LOS path gain measurements collected in the same corridor at 28 GHz (experimental details in [22]) are likewise compared against prediction (19) in Fig. 2. Including wall roughness (second term in (15)) reduces mean error from 3 dB to 0.3 dB, leading to RMS error of 4.1 dB.

Similar agreement is found against 3.5 GHz LOS measurements collected [23] in an urban canyon in Fig. 3, with 1.9 dB RMS error using coherent ground reflection (18) and 2.5 dB RMS error for "incoherent" ground reflection (19).

The observed power and predictions in the unobstructed canyon cases here are above free space, with additional power attributed to reflections from canyon walls.

Extension to the case of transmitter and/or receiver being off-center of the canyon can be obtained by modifying image source locations in (6) and grazing angles (7), as appropriate. The resulting average power expression is more complex yet is found to be insensitive to exact displacement of the antennas and is thus numerically close to (18) and (19), particularly for $r \gg w$. An interesting exception is the case of an antenna placed within a wavelength from a wall, where incoherent power sum (10) is inappropriate as coherent reflection from the near wall is needed to assure the boundary condition is satisfied, requiring coherent sum (5). For very short ranges, where $r < w$, the sum (10) is dominated by the free space term (4), although (19), derived for $r \gg w$, is within 2 dB of that.

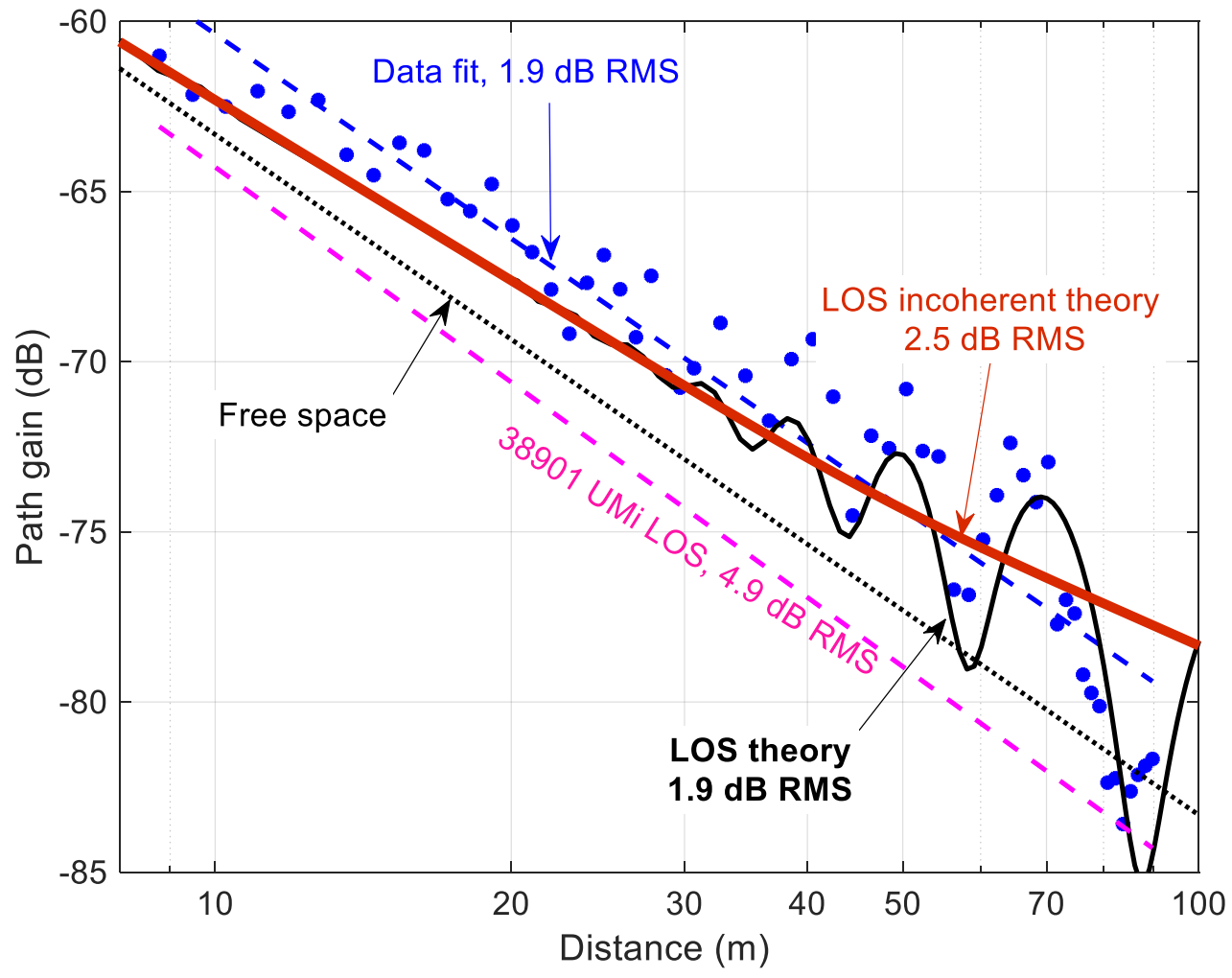

Fig. 3. Path gain in an outdoor LOS urban canyon, 3.5 GHz, 8.6 m-wide street. LOS theory is (18), LOS "incoherent" is (19). Wall loss $L$ (15) parameters: $n_{\text{eff}} = 2.2$, $A$=0.1m, $p_1 = 0.85$, $p_2 = 0.15$, $1/\mu_1 = 0.33$ m, $1/\mu_2 = 2$ m.

This problem was treated in [21][23] using an explicit sum of modes in the corridor/canyon. While accurate, the sum-of-modes approach does not scale well with frequency, as a 30 m-wide street supports some 6000 modes at 28 GHz. Expressions derived here are much more efficient to compute and allow immediate insight into dependence on range and material parameters.

## IV. Penetration Through Foliage in a Suburban Street

### A. *Outdoor terminal on street with trees*

We consider propagation between a base station placed in a region free from clutter (close to middle of a street) and a terminal placed behind foliage. The case arises in suburban Fixed Wireless Access (FWA) where lamppost-mounted base stations communicate with terminals mounted on exterior of buildings, often behind foliage, such as trees and hedges, illustrated in Fig. 4.

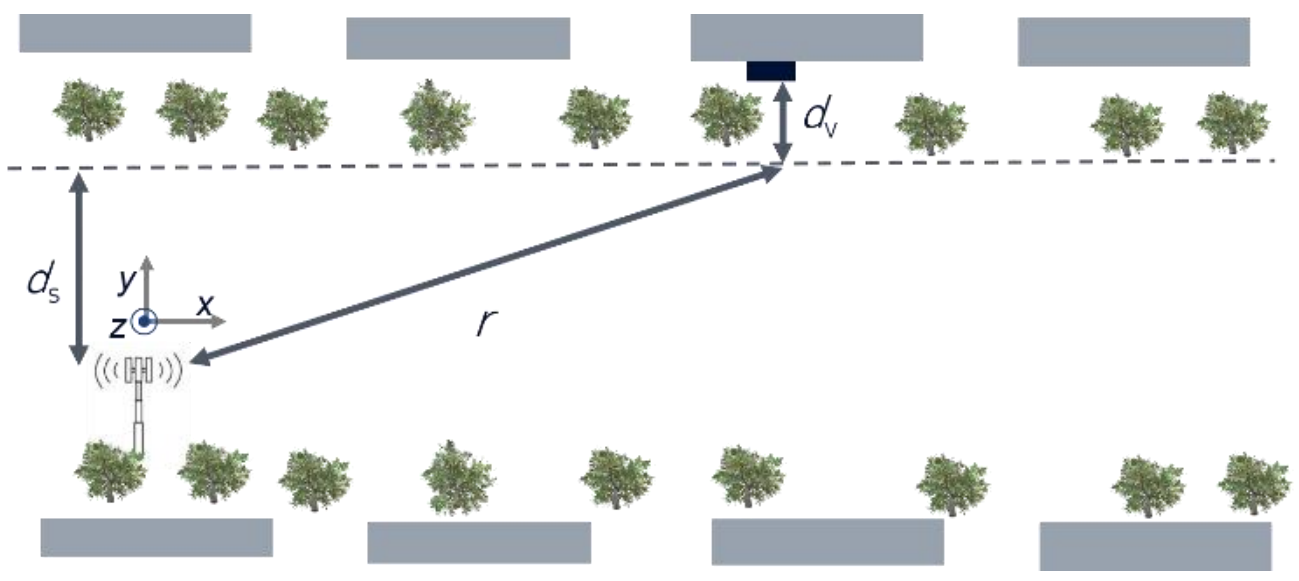

Fig. 4. Suburban link (top view) illustration for a lamppost base (left) and a terminal on house exterior wall, behind vegetation.

Path gain between a source in free space and a terminal at depth $d_v$ in diffuse clutter (here foliage) of absorption $\kappa_v$ (Nepers/m) is derived in Appendix II. Reflections from buildings across the street are neglected due to foliage absorption (~2 dB/m at 28 GHz). Scattering from across-the-street foliage is also neglected. Setting $T_{eff} = 1$ in (63) (no wall between vegetation and open space), and adding power reflected from the ground and the building wall near terminal, the average path gain for suburban street case is given by:

$$P \approx \frac{\lambda^2 d_s^2 e^{-\kappa_v d_v}}{8\pi^2 r^4}\left(1+\left|\Gamma_g\right|^2\right)\left(1+\left|\Gamma_w\right|^2\right) \qquad (20)$$

where $\left|\Gamma_g\right|^2$ and $\left|\Gamma_w\right|^2$ are the ground and wall power reflection coefficients, respectively, $d_s$ is the separation between the source and the nearest point on foliage boundary (Fig. 4), $r = \sqrt{x^2 + (z_s - z)^2 + d_s^2}$ is the distance from source to center of "hot" region on boundary, with base-terminal height difference $z_s - z$. For $r \gg d_v$, $r \approx |\mathbf{r}_s - \mathbf{r}|$, the effective range to terminal. The distance exponent is 4.

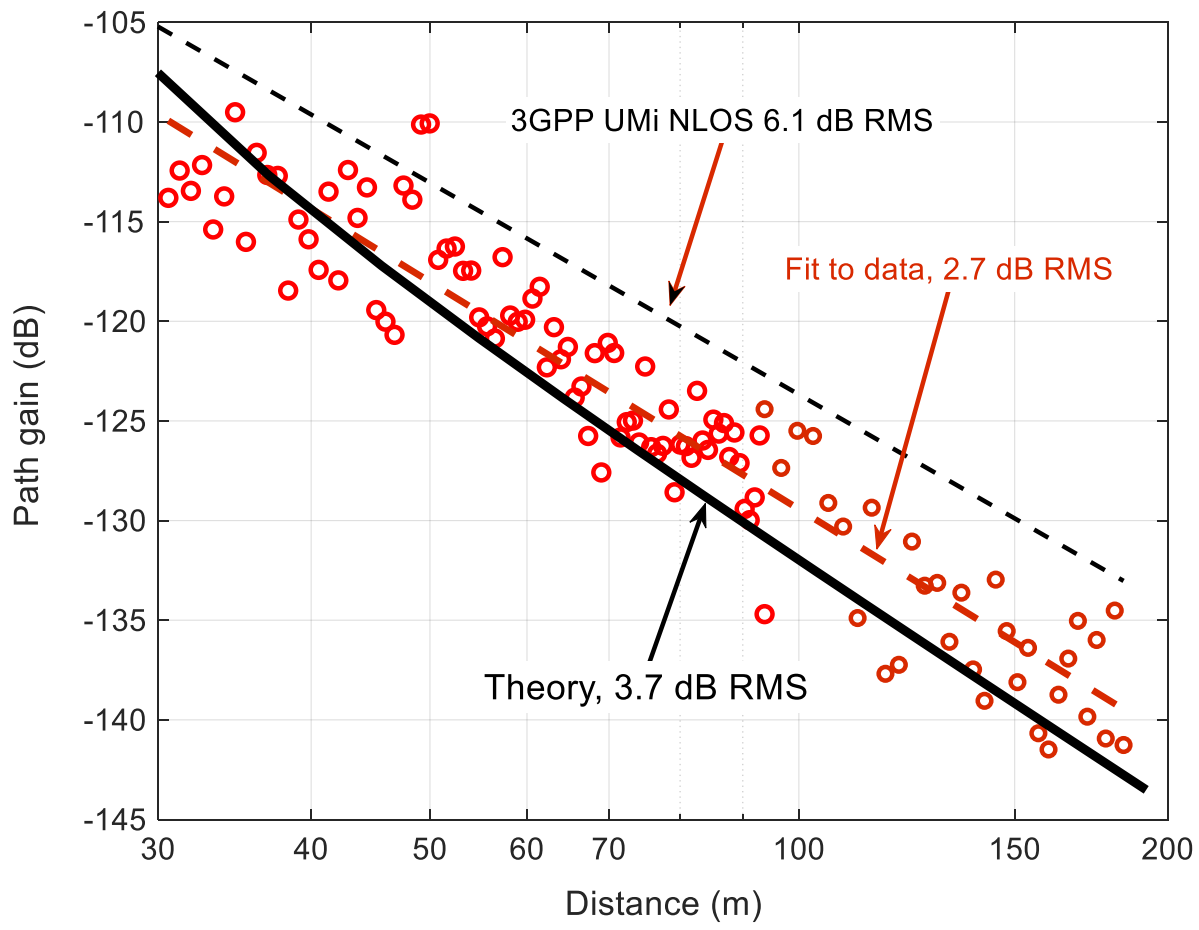

Fig. 5. Suburban street path gain, with 'lamppost' base at 3 m height and terminal at 1 m height on the same street behind 10 m of foliage, on house exterior. Theory is equation (20). $d_s$=20 m, $d_v$=10 m.

Over 1000 NLOS path gain measurements were collected at 28 GHz for the FWA scenario on 6 streets in a NJ suburb [16]. Here we are interested in the same-street scenario, with lamppost node ('base station') and a 'terminal' next to exterior house wall, behind vegetation. In other words, foliage is what changes this arrangement from LOS to NLOS. Average vegetation depth $d_v$ in (20), was found from publicly available satellite views for each measured street, varying from street to street from 3.1 m to 10.7 m. Fig. 5 shows a sample path gain data set plotted vs. distance for a street with about 10 m-thick vegetation layer (trees and bushes) separating the house from the street. Path gain predicted by (20) is found to give 3.7 dB RMS error, only slightly worse than the 2.7 dB RMS deviation from a linear fit to the data and better than the 6.1 dB RMS for the 3GPP UMi NLOS model [3]. Using a street-specific foliage depth in (20) for all 6 streets (~1000 links), results in RMS error of 5.5 dB, as compared to 6.2 dB RMS deviation obtained from a linear fit to the entire data set, and 6.3 dB RMS from 3GPP UMi NLOS, which happens to work well for suburban FWA measurements with dense foliage [16]. An early version of this result was presented in [24].

## B. Indoor terminal on a street with trees

The above formula generalizes directly to the case of a terminal placed inside the building, illustrated in Fig. 6.

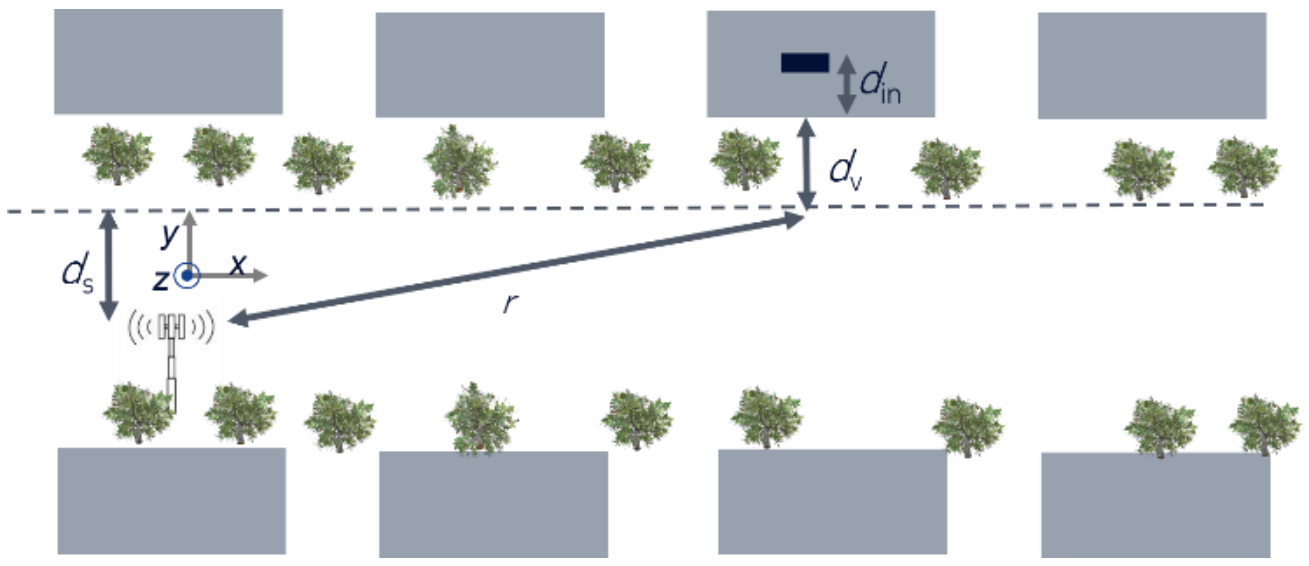

Fig. 6. Suburban link (top view) illustration for a lamppost base (left) and an indoor terminal.

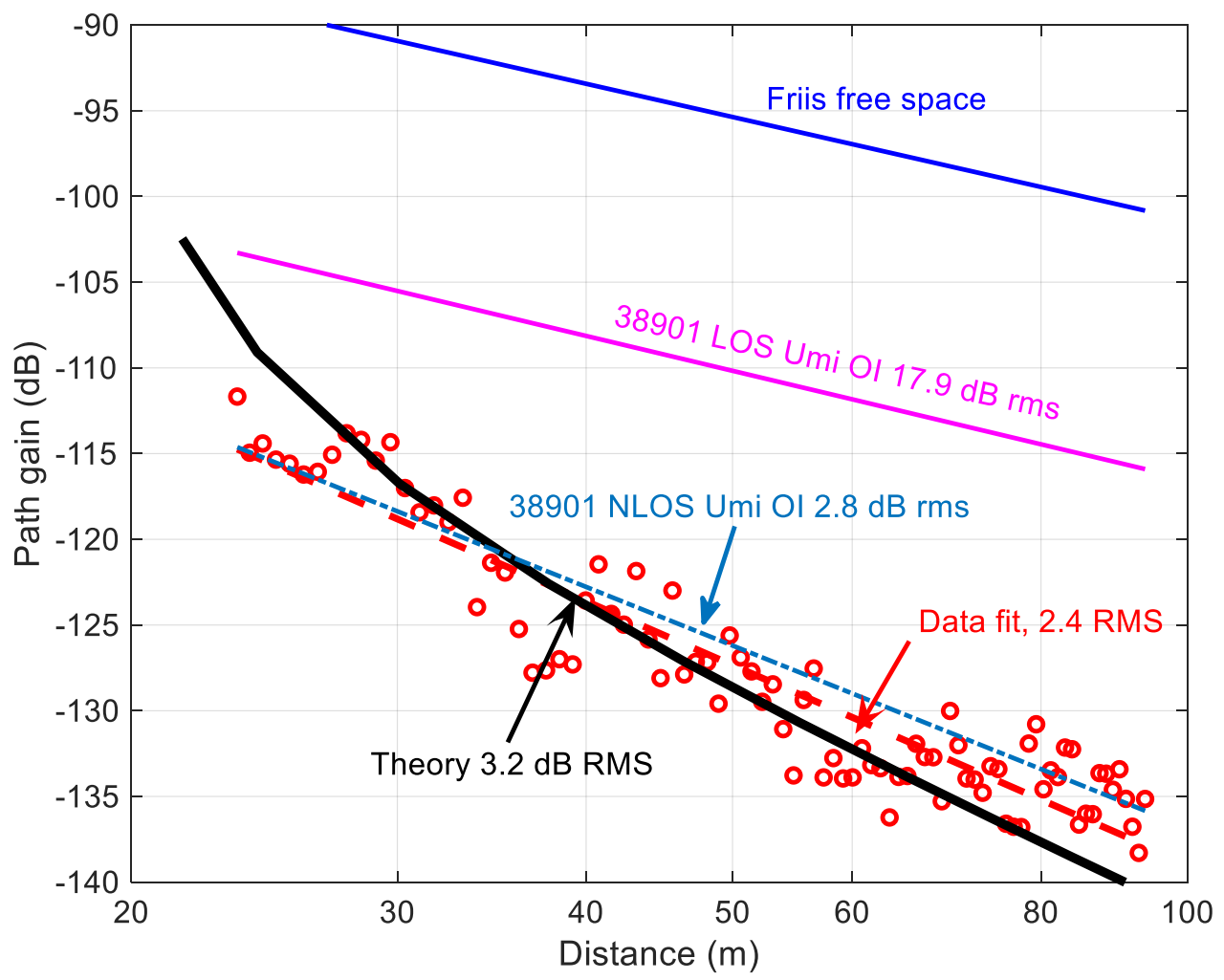

Fig. 7. Outdoor-indoor suburban path gain, with 'lamppost' base at 3 m height and terminal on the same street behind 10 m of foliage, 1.5 m inside the house. Theory is equation (21). $d_s$=20 m, $d_v$=10 m, $d_{in}$=1 m, $T_{eff}$=0.1.

Placement of the terminal inside introduces additional losses to (20): wall penetration loss $T_{eff}$ and loss $e^{-\kappa_{in} d_{in}}$ due to scatter and absorption from indoor clutter, leading to

$$P_{OI} \approx \frac{\lambda^2 d_s^2 e^{-\kappa_v d_v} T_{eff} e^{-\kappa_{in} d_{in}}}{8\pi^2 r^4}\left(1+\left|\Gamma_g\right|^2\right)\left(1+\left|\Gamma_w\right|^2\right) \qquad (21)$$

where $d_s$ is the standoff distance between the source and the nearest point on foliage boundary and $r = \sqrt{x^2 + (z_s - z)^2 + d_s^2}$ is the distance from source to center of "hot" region on boundary, with base-terminal height difference $z_s - z$. $\left|\Gamma_g\right|^2$ accounts for power reflected from the ground, and $\left|\Gamma_w\right|^2$ for power reflected from the back wall of a building. The latter may be approximated either by its maximum value of 1 or an average over all angles between 0 and 90°, to account for wide spectrum illumination in scattering indoor environments. As marked in Fig. 6, $d_v$ is the depth of vegetation, $d_{in}$ is the depth of the terminal indoors. Representative intrinsic material losses in vegetation and interior space at 28 GHz are $\kappa_v$ ~0.38 Nep/m (linearly interpolated in frequency between 0.4 Nep/m at 35 GHz [25] and 0.07 Nep/m at 2 GHz [26]) and $\kappa_{in}$ ~0.18 Nep/m [27], respectively. For $r \gg d_v + d_{in}$, $r \approx |\mathbf{r}_s - \mathbf{r}|$, the effective range to receiver. In the usual case of $\sqrt{x^2 + (z_s - z)^2} \gg d_s$, the distance exponent is 4.

Predictions (21) are compared against measurements in Fig. 7. The path gain data, reported and characterized empirically in [28], was collected inside a suburban house with 2 cm-thick plywood walls with calculated penetration loss $T_{eff} = 0.1$ (mostly determined by 10% of wall area occupied by plain glass windows). Theory (21) has 3.2 dB RMS error, compared to 2.4 dB RMS linear fit deviation.

## V. Above Clutter Base and Below Clutter Terminal

The propagation in the NLOS case of a macro cellular base mounted above rooftops and a below clutter outdoor terminal can also be treated as that between a source in free space and a receiver in a diffusely scattering medium, as in Appendix II. The idealized geometry is illustrated in Fig. 8.

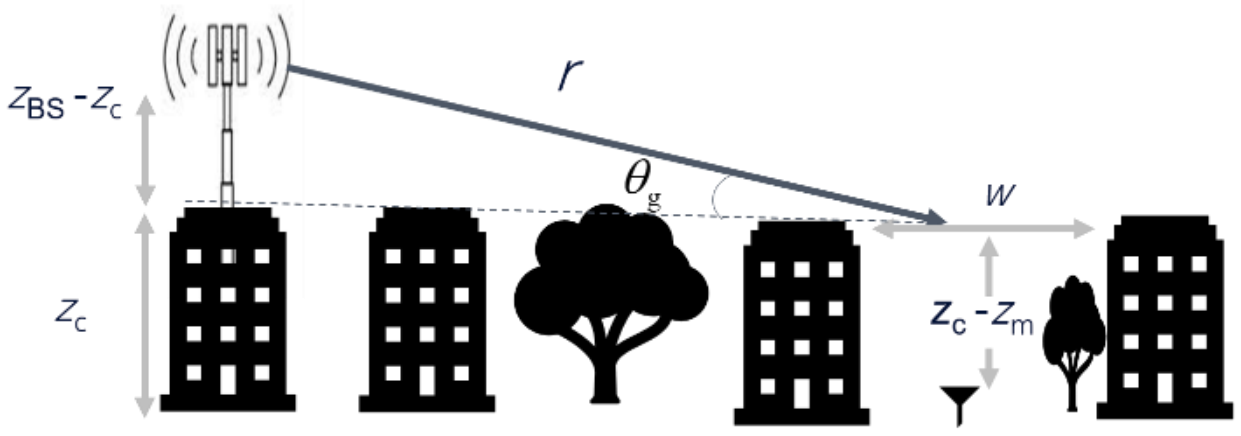

Fig. 8. Rooftop macro to NLOS mobile terminal geometry, with street width *w*, clutter height $z_c$, base height $z_{BS}$, mobile height $z_m$.

The path gain is given by adaptation of (63), adding power reflected from the ground near terminal:

$$P_{\text{over-top}} = \frac{\lambda^2 (z_{BS} - z_c)^2 e^{-\kappa_v |z_c - z_m|} T_{eff} \left(1 + |\Gamma_g|^2\right)}{8\pi^2 r^4} \quad (22)$$

Here the source is $d_s = z_{BS} - z_c$ meters above clutter (local roof tops or tree tops), at range $r = \sqrt{x^2 + (z_{BS} - z_c)^2}$ from the clutter top immediately above the terminal, $T_{eff} = (2/\pi)\tan^{-1}(w/2|z_c - z_m|)$ from middle line of (64) (no material boundary above the street containing the terminal, so $T = 1$), $w$ is the street width and $d = z_c - z_m$ is the "depth" of the terminal below clutter height $z_c$, essentially equal to building height for street level terminals. In the absence of vegetation, absorption $\kappa_v = 0$.

For wide streets, $w \gg (z_c - z_m)$ or rural areas where street width is not well defined, $\tan^{-1}(0.5w/(z_c - z_m)) \to \pi/2$ and (22) simplifies to

$$P_{\text{over-top}} = \frac{\lambda^2 (z_{BS} - z_c)^2 e^{-\kappa_v |z_c - z_m|} \left(1 + |\Gamma_g|^2\right)}{8\pi^2 r^4} \quad (23)$$

For rural areas, in addition to the "over-the-top" path gain (23), there is also a direct path through vegetation, important at short ranges. In such conditions, (23) generalizes to:

$$P_{\text{rural}} = \left(\frac{\lambda}{4\pi r}\right)^2 e^{-\kappa_v r_v} + \frac{\lambda^2 (z_{BS} - z_c)^2 e^{-\kappa_v |z_c - z_m|} \left(1 + |\Gamma_g|^2\right)}{8\pi^2 r^4} \quad (24)$$

where $r_v$ is the part of the direct path going through vegetation. For dense trees of height $z_c = z_{tree}$ less than base antenna height $z_{BS}$:

$$r_v = r \frac{z_c - z_m}{z_{BS} - z_m} \quad (25)$$

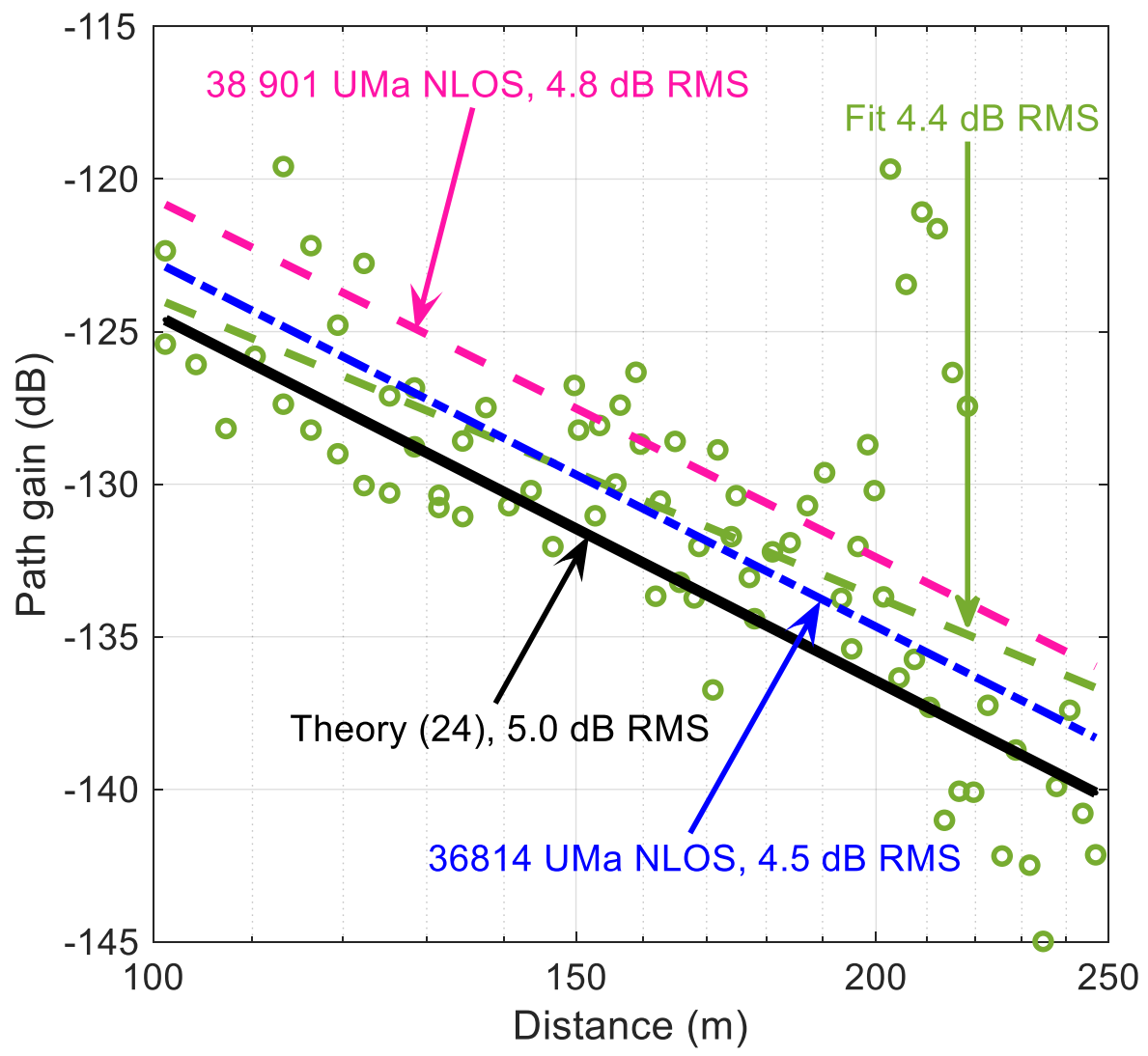

Fig. 9. Measured and predicted path gain on densely vegetated urban street, compared against over-top (24) and 3GPP urban macro models. $z_{BS}$= 14 m, $z_c$=10 m, $z_m$=1.5 m.

Path gain (22) has distance exponent of 4. In conventional macro-cellular deployment the base station is placed above clutter to cover large ranges, so that horizontal range $x \gg (z_{BS} - z_c)$, and the total range $r \approx x$. Equation (24) is a corrected version of [30].

Measurements collected [17] at 28 GHz from a rooftop base to a terminal under the tree canopy of a densely vegetated street are compared in Fig. 9 to (24) and to the 3GPP 36.814 recommendation for urban macrocells [29] (with dependence on building height $z_b$ (here same as clutter height $z_c$) and antenna heights $z_{BS}$ and $z_m$, unlike simplified [3]):

$$\begin{aligned} PL_{\text{UMa_NLOS}} &= 161.04 - 7.1\log_{10}(w) + 7.5\log_{10}(z_{\text{b}}) \\ &- (24.37 - 3.7(z_{\text{b}}/z_{\text{BS}})^2)\log_{10}(z_{\text{BS}}) \\ &+ (43.42 - 3.1\log_{10}(z_{\text{BS}}))(\log_{10}(d_{\text{3D}}) - 3) \\ &+ 20\log_{10}(f_c) - (3.2(\log_{10}(11.75 z_{\text{m}}))^2 - 4.97) \end{aligned} \quad (26)$$

as well as 38.901 UMa NLOS [3]. All predictions are close to each other: 3GPP 36.814 UMa NLOS has 4.5 dB RMS error, 38.901 UMa NLOS 4.8 dB and 5.0 dB RMS for (24).

## VI. "OUTDOOR-INDOOR" COVERAGE FROM STREET CANYONS INTO BUILDINGS AND FROM CORRIDORS INTO ROOMS

Here we are interested in path gain between a terminal inside a building due to a base in a street canyon, say on a lamppost or rooftop overlooking the street, illustrated in Fig. 10. The case is distinct from the suburban propagation examined in Sec. IV in that here foliage is very sparse or absent, allowing reflections from building walls. Very similar modeling applies to a channel between a terminal in a room and a base in a corridor.

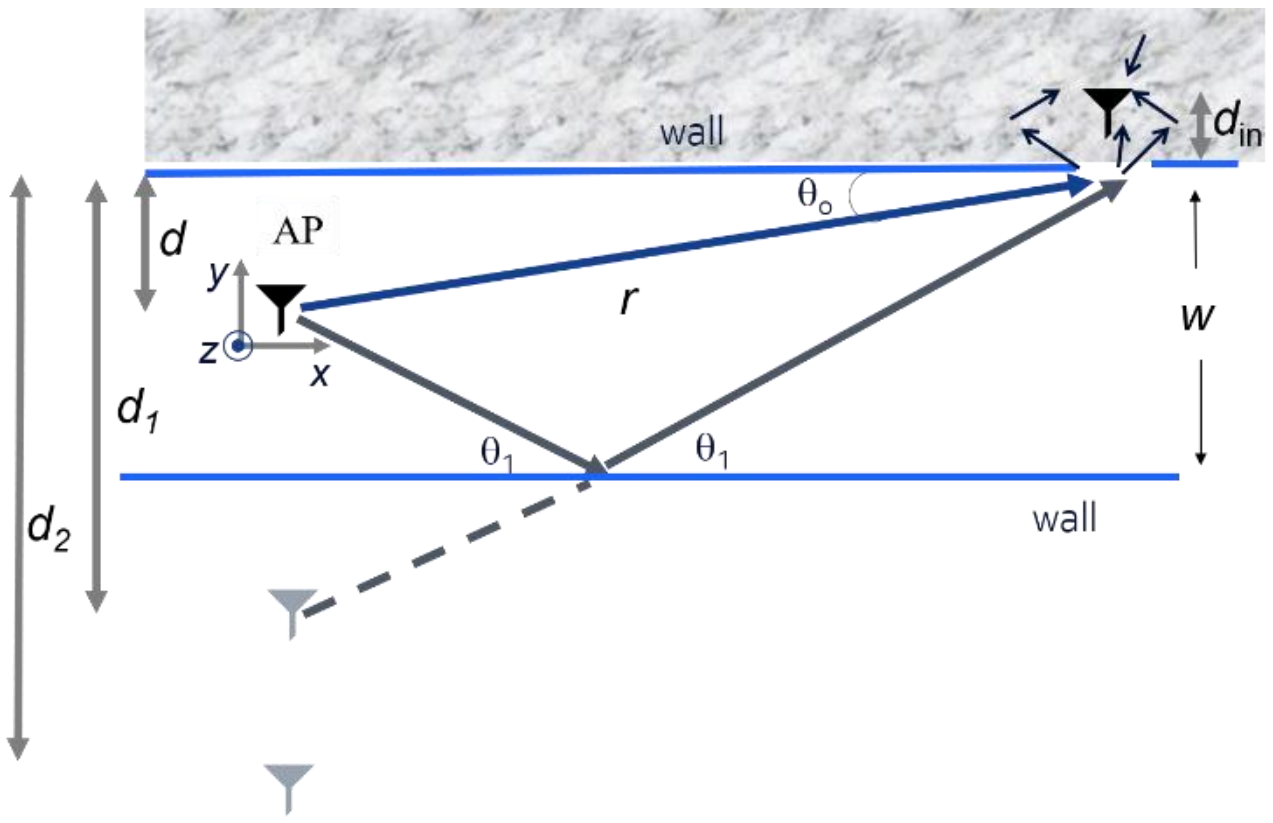

Fig. 10. Top view of the "outdoor-indoor" geometry in a street canyon or indoor corridor-room link. AP is in the canyon, bound by (blue) walls, indoor terminal in cluttered interior at the top.

The field in the street canyon is modeled here as due to multiple reflections from the canyon walls. The "waveguide" field is illuminating the exterior wall of a building containing the terminal. The building interior consists of walls and furniture, modeled here as diffusely scattering medium with indoor absorption parameter $\kappa_{\text{in}}$ (Nep/m), as was done for indoor propagation in [21][23][27] and for vegetation in Sec. IV. The received power due to a direct path from the source with a "standoff distance" $d$ from the building wall (Fig. 10) has been derived in Appendix II as:

$$P_0 \approx \frac{\lambda^2 d^2 e^{-\kappa_{\text{in}} d_{\text{in}}} T_{\text{eff}} \left(1+|\Gamma_{\text{g}}|^2\right)\left(1+|\Gamma_{\text{w}}|^2\right)}{8\pi^2 r^4} \quad (27)$$

where the terminal is placed at depth $d_{\text{in}}$ inside the building, $T_{\text{eff}}$ is the effective (power) transmission coefficient (64) or (65) through the exterior wall and $r$ is the range from a source in the canyon to the center of the "hot" wall on the building exterior, closest to the terminal. The total power reaching the interior terminal is a sum of direct field power (27) and an infinite set of reflected fields from canyon walls:

$$P \approx \frac{\lambda^2 e^{-\kappa_{\text{in}} d_{\text{in}}} T_{\text{eff}} \left(1+|\Gamma_{\text{g}}|^2\right)\left(1+|\Gamma_{\text{w}}|^2\right)}{8\pi^2 r^4} \sum_{m=0}^{\infty} d_m^2 |\Gamma_m|^{2m} \quad (28)$$

where the "standoff" distance of a source image corresponding to $m$ reflections is:

$$\begin{aligned} d_m &= mw + d,\ m = 0, 2, 4... \\ d_m &= mw + w - d,\ m = 1, 3, 5... \end{aligned} \quad (29)$$

Using approximation (14) for reflection coefficient, (28) becomes

$$P \approx \frac{\lambda^2 e^{-\kappa_{\text{in}} d_{\text{in}}} T_{\text{eff}} \left(1+|\Gamma_{\text{g}}|^2\right)\left(1+|\Gamma_{\text{w}}|^2\right)}{8\pi^2 r^4} \times \left\{ \begin{aligned} &\sum_{m=0,2,4}^{\infty} (mw+d)^2 e^{-Lm(m+d/w)w/r} \\ &+ \sum_{m=1,3,5}^{\infty} (mw+w-d)^2 e^{-Lm(m+1-d/w)w/r} \end{aligned} \right\} \quad (30)$$

For $r \gg Lw$ (low wall reflection loss and at grazing incidence), the sums in (30) are dominated by terms for which $m \gg 1 > d/w$. This allows approximating (30) as

$$P \approx \frac{\lambda^2 T_{\text{eff}} \left(1+|\Gamma_{\text{g}}|^2\right)\left(1+|\Gamma_{\text{w}}|^2\right) e^{-\kappa_{\text{in}} d_{\text{in}}}}{8\pi^2 r^4} \sum_{m=0,1,2...}^{\infty} m^2 w^2 e^{-Lm^2 w/r} \quad (31)$$

Similar to the LOS case in Sec. II, the sum in (31) is now approximated as an integral over a continuous variable $s$:

$$P \approx \frac{\lambda^2 e^{-\kappa_{\text{in}} d_{\text{in}}} T_{\text{eff}} \left(1+|\Gamma_{\text{g}}|^2\right)\left(1+|\Gamma_{\text{w}}|^2\right) w^2}{8\pi^2 r^4} \int_0^{\infty} s^2 e^{-Ls^2 w/r} ds \quad (32)$$

Evaluating the integral leads to the final expression for the average path gain in the outdoor-indoor canyon scenario:

$$P_{\text{canyon, out-in}} \approx \frac{\lambda^2 T_{\text{eff}} \left(1+|\Gamma_{\text{g}}|^2\right)\left(1+|\Gamma_{\text{w}}|^2\right) e^{-\kappa_{\text{in}} d_{\text{in}}}}{32\pi^{1.5}} \frac{\sqrt{w}}{L^{1.5} r^{2.5}} \quad (33)$$

dependent on wavelength $\lambda$, window/wall power transmission coefficient $T_{\text{eff}}$, indoor absorption $\kappa_{\text{in}}$, indoor terminal depth $d_{\text{in}}$, street width $w$, and loss parameter $L$ (15), in turn dependent on wall material and roughness. Back wall reflection coefficient is approximated as $|\Gamma_{\text{w}}|^2 \approx 1$ for exterior wall confining the radiation within the building. Long range received power decreases with distance $r$ with an exponent of 2.5, which is independent of wall properties and street width.

Narrowband received power measurements were collected [23] at 3.5 GHz in a street on the campus of Universidad Técnica Federico Santa María (UTFSM), Valparaiso, Chile. The 8.6 m street canyon was lined by concrete buildings, with clear glass windows occupying about 30% of the wall area. The windows were separated by concrete pillars, with the resulting concrete-glass surface having 0.2 m depth of corrugation. A 3.5 GHz tone was emitted from a 10.2 dBi transmit patch antenna, placed 0.5 m from a wall at 5 m height, aimed "down the street". The receiver was a 2.4 dBi "whip" antenna, placed at a height of 7.7 m inside the building at depths varying from 1 m to 6 m. The comparison of locally averaged receive power obtained from measurements, prediction (33), and free space loss are illustrated in Fig. 11. Measurements collected with transmitter on different sides of the street are distinguished: marked blue for same side of the street as the building containing the receiver and red for opposite side of the street. Measured average power was obtained by averaging the instantaneous received power over 60 local displacements of an omnidirectional receive antenna as it is rotated over a circle of 0.4 m radius. Path gain formula (33) has 2.8 dB RMS error. The 3GPP UMi LOS O2I [3] model also does well with 3.3 dB RMSE.

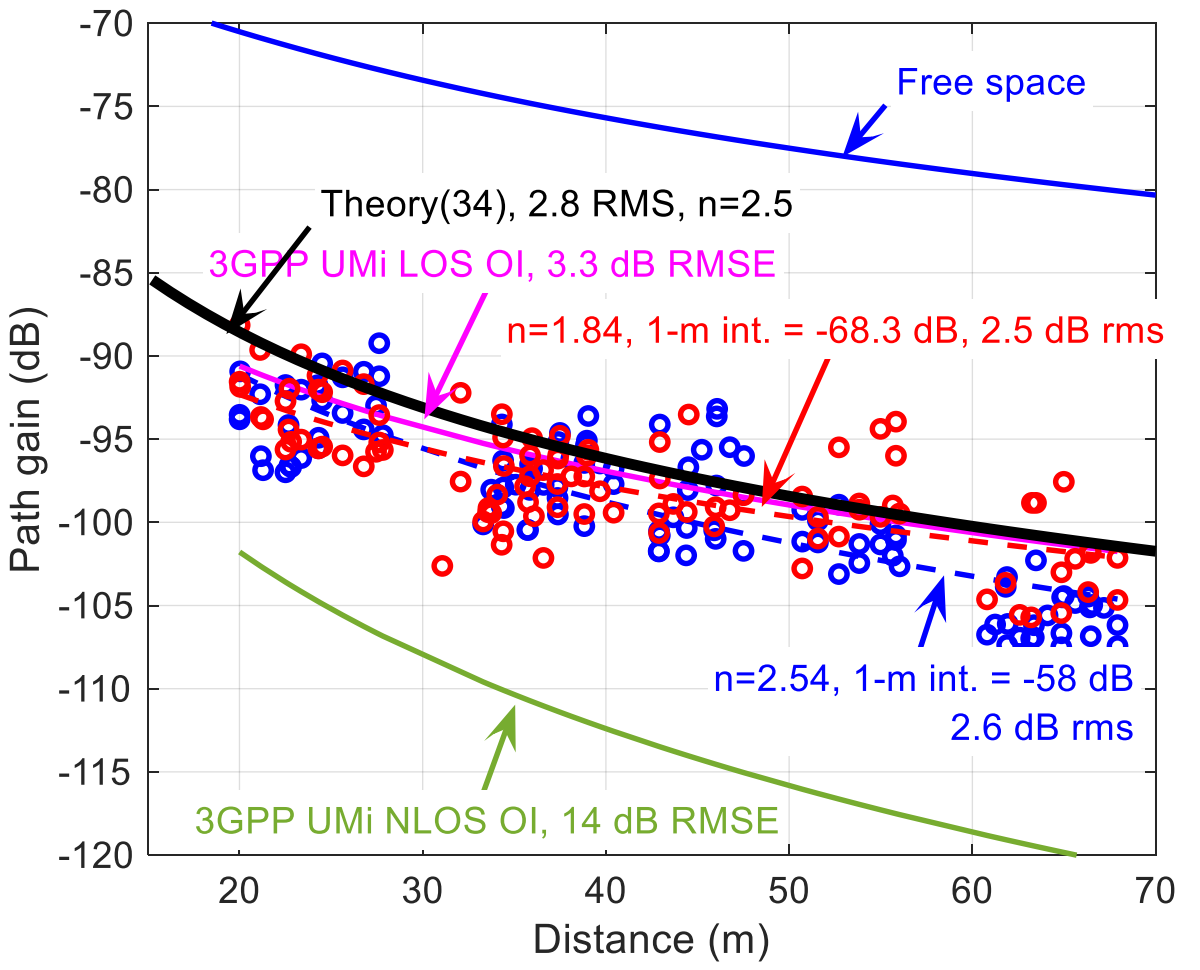

Fig. 11. Measured and predicted signal power for outdoor-indoor case in a street waveguide at 3.5 GHz. Tx antenna patch mounted close to the wall either opposite (red) or same side (blue) of the street as the building containing the receiver. Linear fits are dashed, theory (33) is solid. Wall loss $L$ (15) parameters: $n_{\mathrm{eff}} = 2.2$, $A$=0.1m, $p_1 = 0.85$, $p_2 = 0.15$, $1/\mu_1 = 0.33$ m, $1/\mu_2 = 2$ m. $T_{\mathrm{eff}}$=0.37, $w$=8.6m.

The corresponding WINNER Outdoor-indoor model [4], based on [20], predicted the path gain with RMS errors of 3.9 dB for the transmitter on the opposite side and 6.7 dB RMS error for the same side.

The fit lines to the same and opposite side data are within 2 dB of each other, despite a large difference in the angle of incidence on the wall of the building containing the terminal. The small path gain difference may be explained by recognizing that reflections from the canyon walls occur at steeper angles of incidence than direct illumination. Their inclusion in (33) provides for close agreement with measurements whether the base in the street is close or far from the face of the building.

Predicted path gain (33) also applies to the morphologically similar NLOS indoor case of a base in the corridor and a terminal inside a room adjacent to that corridor. Path gain (33) is compared against 74 link measurements (details in [21]) at 2 GHz and 452 links (details in [22]) at 28 GHz in Fig. 12, with RMS errors of 3.9 dB for both frequencies. $T_{\mathrm{eff}}$ is approximated by (65), where the open door plays the role of a fully transparent "window" occupying $p$=0.25 fraction of the wall. 3GPP indoor NLOS model [3] has corresponding RMS errors of 5.7 and 7.5 dB

Measured NLOS excess loss relative to free space increases from 14 dB to 27 dB at 70 m, as the frequency is increased from 2 GHz to 28 GHz. Theoretical model (33) reproduces this scaling with frequency, through increased rough wall scattering loss $L$ (15), dependent on frequency through wavenumber $k = 2\pi f/c$.

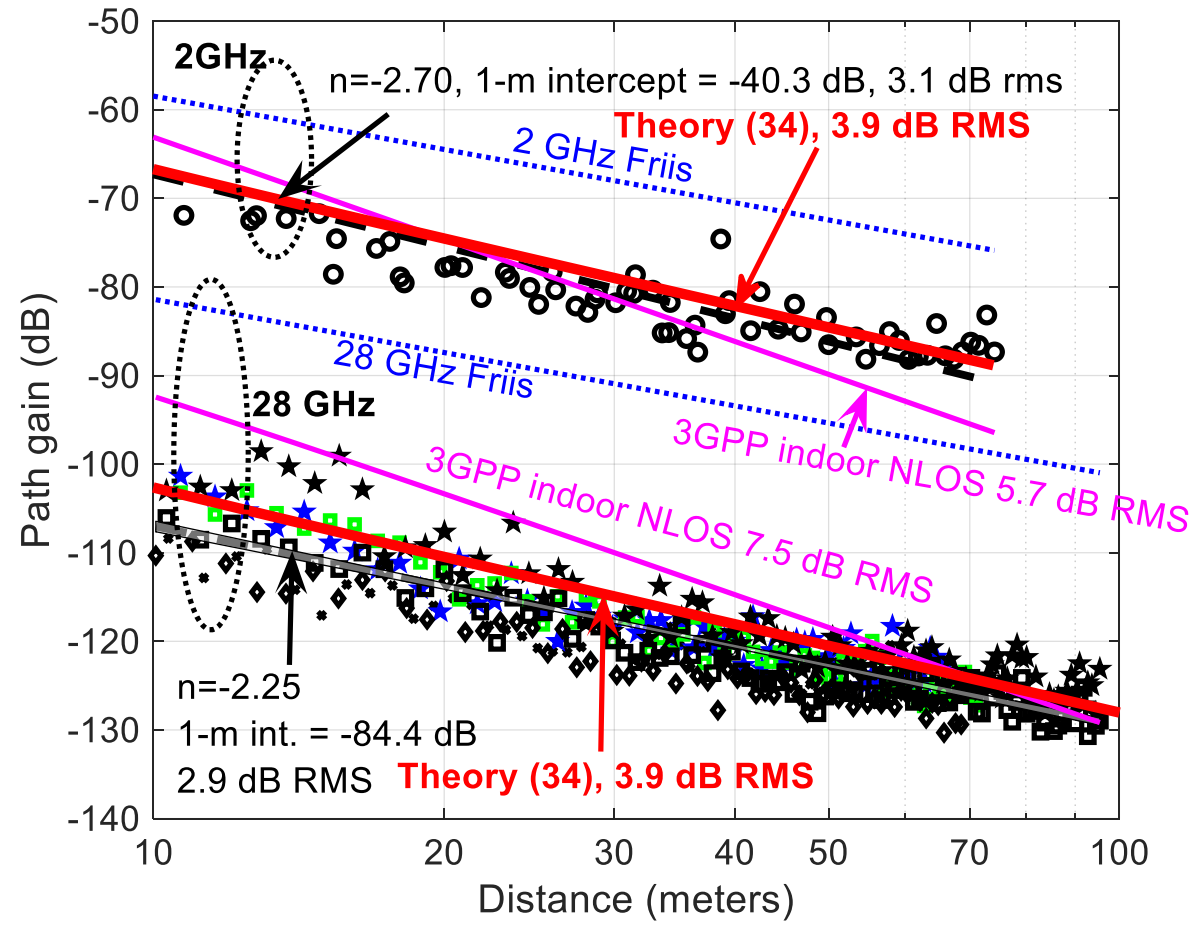

Fig. 12. Path gain in indoor corridor-room NLOS at 2 GHz and 28 GHz. Wall loss $L$ (15) parameters: $n_{\mathrm{eff}} = 1.7$, $A$=0.035m, $p_1 = 0.25$, $p_2 = 0.75$, $1/\mu_1 = 1$ m, $1/\mu_2 = 3$ m, $T_{\mathrm{eff}}$=0.27.

This problem has been analyzed in [21][23] using sum of modes in the canyon waveguide, penetrating into the scattering interior space behind a wall. As mentioned in the LOS case in Sec. II, the sum-of-modes approach, while accurate, is not efficient to compute as frequency increases. Simple expressions derived here are not only efficient to evaluate but also allow insight into dependence on range and material parameters.

## VII. Urban Canyon to Terminal on Cluttered Sidewalk

### A. *Urban sidewalk with trees*

Here we examine the case of a base antenna on a rooftop/lamppost and a terminal on a sidewalk "down the street", illustrated in Fig. 13.

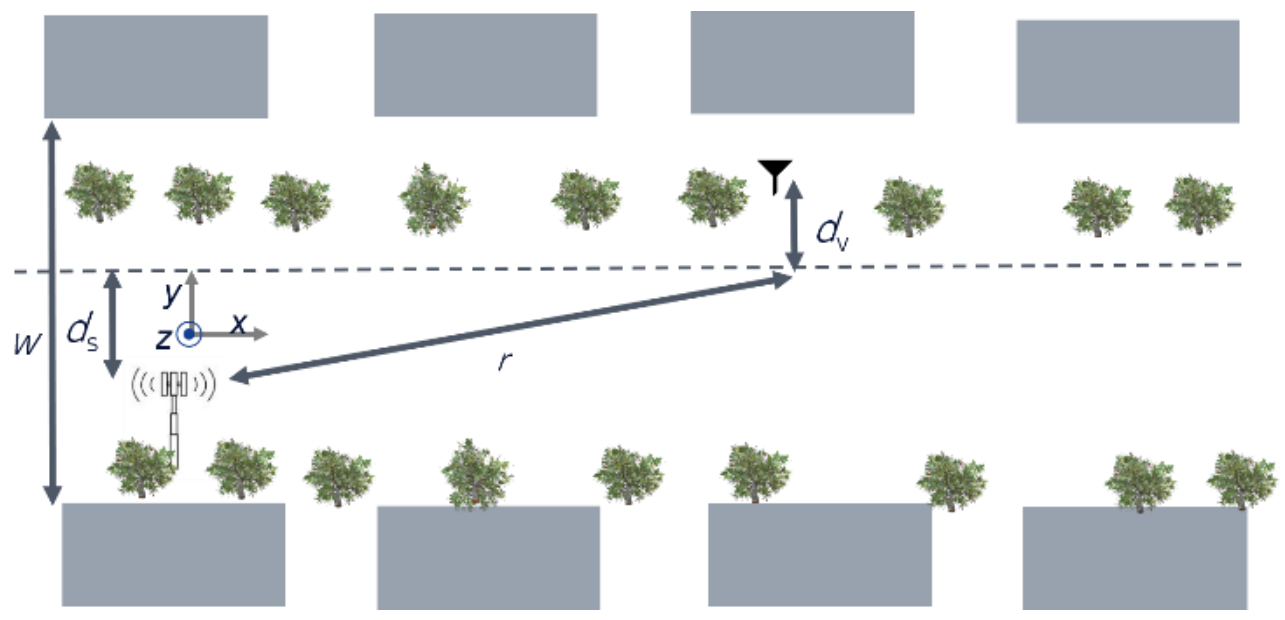

Fig. 13. Lamppost base to terminal on cluttered sidewalk.

This is treated as a generalization of the "outdoor-indoor" guided canyon case in Sec. VI, adding absorption through vegetation to the sum of reflections in (28) and setting $T_{\rm eff}=1$ (no material boundary between street and vegetation):

$$P_{\rm guided}\approx\frac{\lambda^2e^{-\kappa_{\rm v}\rho_{\rm v}d_{\rm v}}\left(1+\left|\Gamma_{\rm g}\right|^2\right)\left(1+\left|\Gamma_{\rm w}\right|^2\right)}{8\pi^2r^4}\sum_{m=0}^{\infty}d_m^2\left|\Gamma_m\right|^{2m}e^{-\kappa_{\rm v}\rho_{\rm v}r_m} \quad (34)$$

where $\rho_{\rm v}r_m$ represents path length through vegetation for $m^{\rm th}$ order reflection path. Length $r_m$ is approximated as

$$r_m\approx\sqrt{r^2+d_m^2}\approx r+\frac{d_m^2}{2r}\approx r+\frac{m^2w^2}{2r} \quad (35)$$

while $\rho_{\rm v}\in[0,1]$ is the tree density on the street, i.e. fraction of canyon volume below base station occupied by trees, for average tree height $z_{\rm tree}$, base height $z_{\rm BS}$, mobile height $z_{\rm m}$, and average tree crown width $w_{\rm tree}$ (e.g. ~ 3-5 m) , and $n_{\rm tree}$ per meter (linear tree density along the street) estimated as

$$\rho_{\rm v}\approx n_{\rm tree}\frac{\left(z_{\rm tree}-z_{\rm m}\right)}{\left(z_{\rm BS}-z_{\rm m}\right)}\frac{2w_{\rm tree}}{w} \quad (36)$$

Accounting for vertical $\left(z_{\rm tree}-z_{\rm m}\right)/\left(z_{\rm BS}-z_{\rm m}\right)$ and horizontal $2w_{\rm tree}/w$ fractions of the canyon, assuming trees on both sides of the street, and $z_{\rm BS}>z_{\rm tree}$, $2w_{\rm tree}<w$. Following the same derivation steps as those leading to (33), leads to:

$$P_{\rm guided}\approx\frac{\lambda^2(1+\left|\Gamma_{\rm g}\right|^2)\left(1+\left|\Gamma_{\rm w}\right|^2\right)e^{-\kappa_{\rm v}\rho_{\rm v}d_{\rm v}}e^{-\kappa_{\rm v}\rho_{\rm v}r}}{32\pi^{1.5}}\frac{\sqrt{w}}{L_1^{1.5}r^{2.5}} \quad (37)$$

where the loss parameter $L$ (15) was modified to include excess absorption suffered by higher order canyon reflections passing through vegetation:

$$L_1=L+\kappa_{\rm v}\rho_{\rm v}w/2 \quad (38)$$

Guided propagation (37) suffers exponential loss due to vegetation through the factor $e^{-\kappa_{\rm v}\rho_{\rm v}r}$. Intrinsic absorption through vegetation at 28 GHz of $\kappa$ =0.38 Nep/m (~2 dB/m) leads to 10 dB of attenuation through a single tree with a 5-m crown, leading to severe attenuation at ranges of interest when even a few trees are present. Notably, path directly illuminating the vegetation near terminal suffers absorption only through depth $d_{\rm v}$ (~ 5 m) and does not depend on range $r$. The propagation is here similar to the suburban case in Sec.IV, essentially down the middle of the street above vehicles and through the sidewalk clutter (mostly vegetation) towards the terminal on a sidewalk. The path gain is then given by (20):

$$P_{\rm unguided}\approx\frac{\lambda^2d_s^2e^{-\kappa_{\rm v}\rho_{\rm v}d_{\rm v}}}{8\pi^2r^4}\left(1+\left|\Gamma_{\rm g}\right|^2\right)\left(1+\left|\Gamma_{\rm w}\right|^2\right) \quad (39)$$

Power represented by (39) corresponds to the lowest order term in the sum (34), avoiding approximation (35). For $d_{\rm s}\ll w$ and base station antennas above trees, next order contribution is from a path reflecting from buildings across the street (bottom of Fig. 13) prior to illuminating the scattering region near terminal. For a base near middle of the street, this is modeled by setting $d_{\rm s}=w$ in (39).

An effective way to combine the low order contribution (39) and the infinite sum (37) is through:

$$P_{\rm canyon,trees}=\max\left[P_{\rm guided},P_{\rm unguided}\right] \quad (40)$$

It was found that when even a few trees are present, (39) dominates, with distance exponent 4. Guided propagation (37) arises only when there are so few trees so as to make loss due to $e^{-\kappa_{\rm v}\rho_{\rm v}r}$ negligible, leading to distance exponent of 2.5.

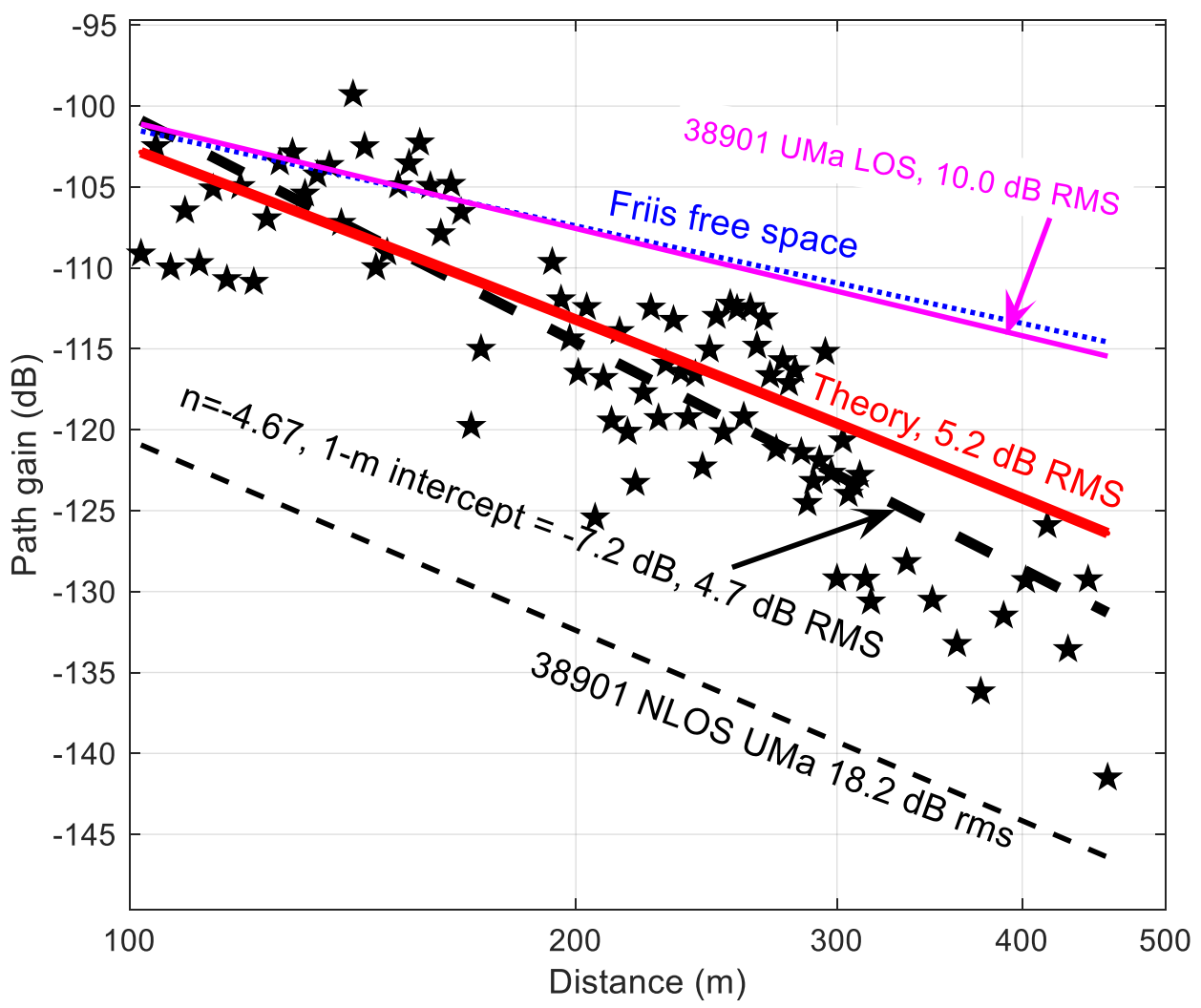

Fig. 14. Path gain at 28 GHz on a Manhattan sidewalk with trees 15 m apart. Theory is (40), dominated by unguided path (39).

Path gain predicted by (40) is compared to path gain measurements collected [17] on two representative Manhattan streets: street with 15 m average tree separation in Fig. 14 and 100 m tree separation in Fig. 15. Measured path gains are seen to be very different: at 400 m, fit to path gain data from a street with denser trees in Fig. 14 is about 7 dB lower than in the case of a street with sparser trees in Fig. 15. For the denser tree case in Fig. 14, equation (40), dominated by (39), predicts the measured path gain with 5.2 dB RMS, as compared to 4.7 dB RMS obtained with a slope-intercept fit to data and 10 dB RMS with 3GPP UMa LOS model.

For the street with sparser trees in Fig. 15, theory (40), now dominated at ranges beyond 200 m by the guided contribution (37), has RMS error of 4.8 dB, as compared to 3.6 dB RMS linear fit deviation and 7.5 dB 3GPP UMa LOS [3] error.

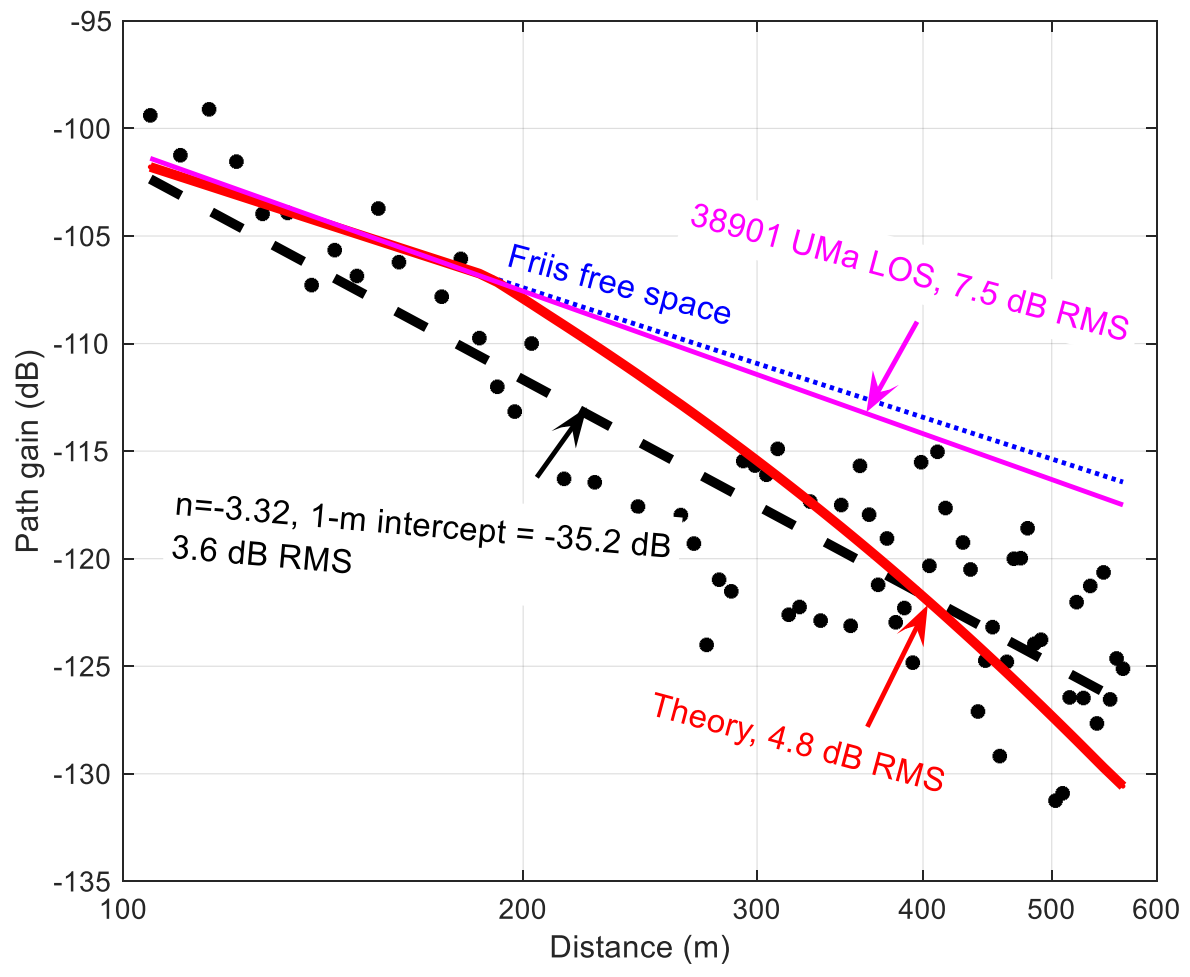

Fig. 15. Rooftop to sidewalk path gain at 28 GHz on a street with trees 100 m apart. Theory is (40), dominated by guided propagation (37), in addition to over-top component (25).

### *B. Diverse urban canyon measurements vs. model*

Path gain from over 800 links measured on 12 streets in Manhattan at 28 GHz [17] is here compared against formulae presented above. In all cases, a rooftop ("BS") antenna was used to record received power from an omnidirectional transmitter 1.5 m above street level, placed on the same street as the rooftop antenna. The joint data set is shown in Fig 16.

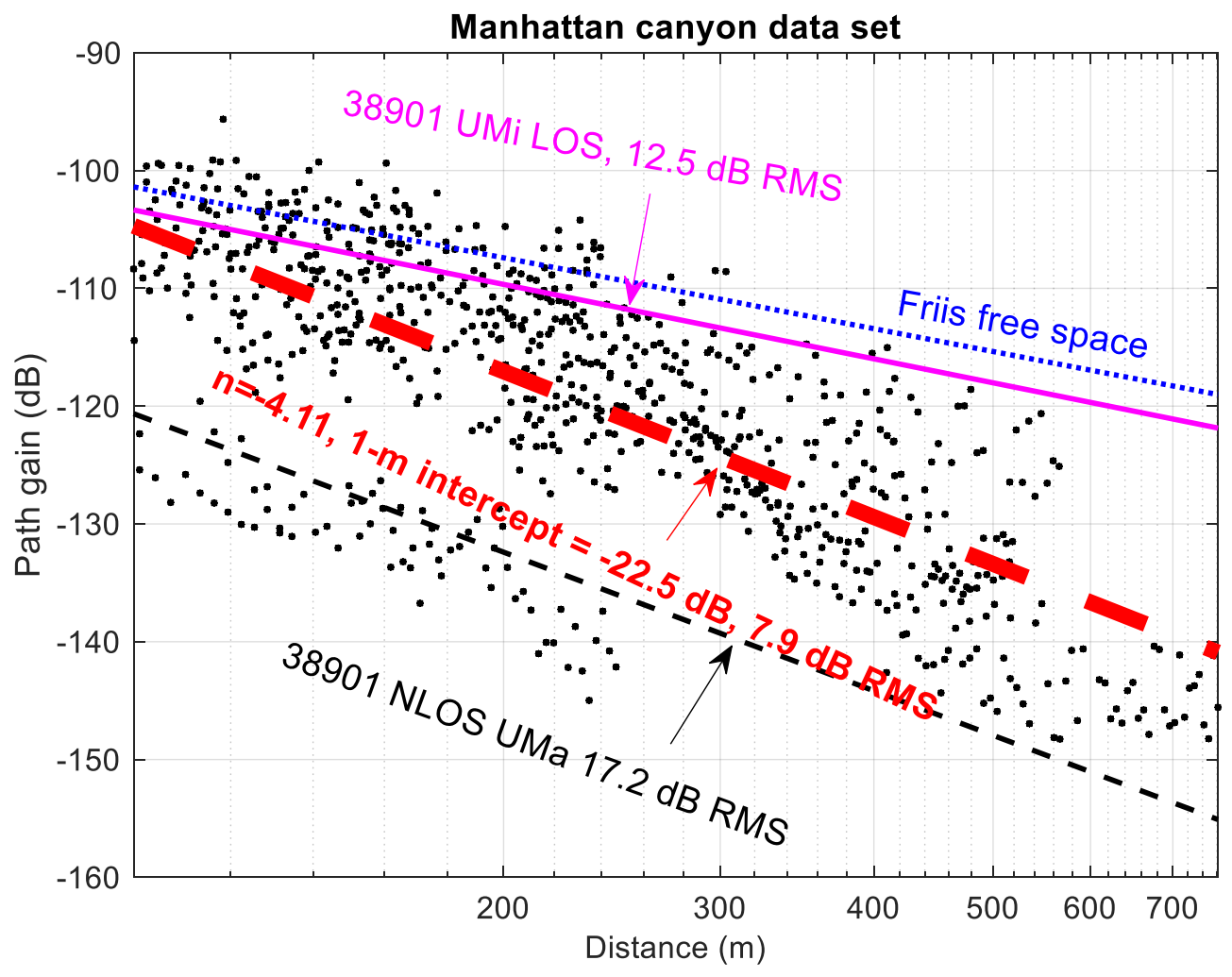

Fig. 16. Measured path gain in 12 streets in Manhattan.

Median measured path gain at similar distance on different streets spans over 30 dB [17], indicating diverse propagation conditions and leading to 7.9 dB RMS goodness of fit spread around the slope-intercept fit. For example, fit to measured path gain at 250 m on a densely vegetated street in Fig. 9 is some 23 dB lower than on a street with sparse trees in Fig. 15.

Various propagation mechanisms presented in previous sections contribute to the overall signal power in urban street canyons. The theoretical model applied consists of the sum of powers for a path penetrating clutter from the side (40), over-top of trees contribution (23) and a direct path, attenuated through clutter:

$$P_{\text{canyon-total}} = P_{\text{canyon,trees}} + P_{\text{over-top}} + P_{\text{direct}} \tag{41}$$

with

$$P_{\text{direct}} = \left(\frac{\lambda}{4\pi r}\right)^2 e^{-\kappa_v r_v} \tag{42}$$

where $r_v$ is the part of direct range passing through vegetation, as estimated from tree locations. Even coarse tree information, such as "10 m average height trees, along both sides, every 20 m" may be used to estimate $r_v$. In the absence of vegetation, pedestrians may become dominant with a scattering cross-section of $\sigma_{\text{ped}} = 1\,\text{m}^2$, area density of about 1 pedestrian/25 m$^2$ area, which, for 2 m height, corresponds to volumetric scattererer density of $\rho_n = 1$ person/50 m$^3$. The intrinsic absorption for such sidewalk environment is then $\kappa_{\text{ped}} = \rho_n \sigma_{\text{ped}} = 0.02$ Nep/m. Similarly, attenuation due to scaffolding, often present on NYC streets, is estimated as $\kappa_{\text{scaff}} = 0.1$ Nep/m.

The composite model (41) is a mixture of components with specific distance exponents, although it is often dominated by a particular phenomenon, depending on base station and vegetation heights and vegetation distribution. For streets with tree canopy covering the entire street width, the 1st term in (41) (penetration from the side) is heavily attenuated and over-top and direct components in (41) dominate. The exponential absorption loss at 28 GHz is very substantial for all but small values of $r_v$ and contribution from (42) was found to be significant only at short ranges on one street (W. 11 Str.) where the first 200 m were free of vegetation, providing for near LOS conditions there. Since each street has its own predicted path gain curve, dependent on local conditions, as in Figs. 9, 14 and 15, the 12 predicted curves are not plotted in Fig. 16 for clarity.

The accuracy of the theoretical formula (41) and 3GPP urban Macro models (38.901 UMa, [3]) is summarized for each street measured, as well the data set overall in Table 1.

For each individual street, the theoretical formulae are seen to be generally more accurate than the 3GPP formulae. A separate fit to data from each street is usually most accurate but unavailable until the data is collected, making predictions unnecessary. Most desired aspect of a model is its ability to predict measurements yet to be made. It may be observed in Table 1 that the theoretical model described in this section is able to take into account general conditions, such as vegetation amounts in the street, giving overall RMS error of 4.7 dB, comparing favorably against 7.9 dB RMS fit to overall data set and 3GPP model RMS errors of 13.9/17.2 dB.

Evaluating (41) at 2 GHz for the same set of streets leads to predicted path gain within 5 dB of the corresponding Friis free space formula at ranges up to 500 m on most streets except ones with particularly high tree density. Much lower foliage absorption $\kappa_v$ at lower frequencies (0.3 dB/m at 2 GHz) [26] makes the attenuated direct path (42) more prominent in (41). This may explain the 3GPP UMa LOS recommendation being close to free space: more reasonable at cellular frequencies than in the mm wave band.

*Table 1. RMS errors (dB) of path gain models on urban sidewalks. Environmental parameters for urban canyon measurements: heights of BS, clutter (tree/scaffolding), street width w, veg. depth $d_v$ and fraction $n_{tree}$ of street length filled by trees. Wall loss L (15) parameters:* $n_{eff} = 2.2$, *A=0.01m*, $p_1 = 0.85$, $p_2 = 0.15$, $1/\mu_1 = 0.33$ m, $1/\mu_2 = 2$ m.

| Street parameters | Data Fit | 38.901 UMa LOS | 38.901 UMa NLOS | Theory (41) |
|---|---|---|---|---|
| 120 Str. NW. (winter), *w*=35m, $h_{BS}$=15m,$h_c$=10m, $d_v$=2m, $n_{tree}$=0.25 | 3.7 | 3.9 | 22.7 | **5.3** |
| 120th N.W. (summer), *w*=35m, $h_{BS}$=15m,$h_c$=10m, $d_v$=2m, $n_{tree}$=0.25 | 5 | 6.3 | 20.2 | **5.3** |
| Amsterdam N.E., *w*=35m, $h_{BS}$=15m,$h_c$=10m, $d_v$=2m, $n_{tree}$=0.25 | 4.1 | 11.9 | 18.8 | **5.2** |
| Amsterdam N.W., *w*=35m, $h_{BS}$=15m,$h_c$=10m, $d_v$=2m, $n_{tree}$=0.25 | 3.2 | 14.7 | 13.9 | **4.6** |
| 120 Str. S.E., *w*=30m, $h_{BS}$=15m,$h_c$=6m, $d_v$=5m, $n_{tree}$=0.25 | 2.6 | 6.5 | 16.1 | **2.6** |
| 120 Str. N.E. , *w*=30m, $h_{BS}$=15m,$h_c$=10m, $d_v$=5m, $n_{tree}$=0.25 | 4.3 | 6.5 | 17.6 | **4.8** |
| Morningside, *w*=30m, $h_{BS}$=14m,$h_c$=10m, $d_v$=5m, $n_{tree}$=0.5 | 4.4 | 25.6 | 4.8 | **5.0** |
| 1st Ave, *w*=44m, $h_{BS}$=22m,$h_c$=10m, $d_v$=5m, $n_{tree}$=0.5 | 4.7 | 10.0 | 18.2 | **5.2** |
| 3rd Ave, *w*=32m, $h_{BS}$=56m,$h_c$=10m, $d_v$=3m, $n_{tree}$=0.05 | 3.8 | 5.8 | 22.9 | **4.5** |
| E_Broadway, *w*=40m, $h_{BS}$=20m,$h_c$=10m, $d_v$=10m, $n_{tree}$=1 | 4 | 10.6 | 16.5 | **4.5** |
| 7th Ave N, *w*=32m, $h_{BS}$=20m,$h_c$=3m, $d_v$=2m, $n_{tree}$=0 | 3.6 | 7.5 | 22.1 | **4.8** |
| W. 11 Str, *w*=20m, $h_{BS}$=20m,$h_c$=10m, $d_v$=3m, $n_{tree}$=0.2 | 4.2 | 22.2 | 11.6 | **3.3** |
| **Overall** | 7.9 | 13.9 | 17.2 | **4.7** |

## VIII. Conclusions

A model for path gain in several canonical and typical scenarios has been derived starting from electromagnetic fundamentals. The model is accurate, with RMS errors under 5 dB, and depends on generally available parameters, such as antenna heights, street width and vegetation amounts. The proposed model adds multipath in power rather than coherently, thus effectively removing multipath fading. Apart from the frequency dependent Friis loss, additional frequency dependence results from wall roughness (through parameter *L*), penetration through materials (transmission coefficient) and through foliage (absorption loss).

Predicted path gain is explicitly derived to be "average" and does not include variation due to multipath fading. Such variation can be added separately, for example by distributing the total predicted power to different multipaths, as in [3]. The model may be used for either site-specific coverage calculations (using local parameters) or generic studies using reasonable default parameters. For urban areas, the model presented here includes propagation down the same street and "true NLOS" over-top propagation for rooftop base antennas. Propagation around-the-corner may be added through a corresponding model, e.g. [17][36].

In all cases analyzed, path gain at long ranges was found to depend on inverse distance with an exponent specific to each environment type:

- 1.5 for LOS canyon and indoor corridors (wall reflections increase received power over free space);
- 4 for urban/rural macro-cellular (above rooftop/treetop to terminal in clutter with small grazing incidence), and for lamppost to terminal behind foliage (i.e. shot from side to terminal embedded in foliage/clutter);
- 2.5 for "outdoor-indoor" scenario for no-foliage urban street canyon and corridor to room indoors (wall reflections provide guiding).

## Appendix 1: Approximation to Reflection Coefficient

Plane wave reflection coefficient from half space with relative index of refraction $n_2$ at grazing angle $\theta_g$ for perpendicular polarization is given by [31]:

$$\Gamma_0 = \frac{\sin\theta_g - \sqrt{n_2^2 - \cos^2\theta_g}}{\sin\theta_g + \sqrt{n_2^2 - \cos^2\theta_g}} \tag{43}$$

At low grazing angles $\theta_g \ll 1$ and $n_2 \gg 1$, (e.g. $n_2 \approx \sqrt{5}$ for concrete) (43) can be approximated by the two lowest order terms of its Taylor series:

$$\Gamma_0 \approx -1 + \frac{2}{n_2}\theta_g \approx -e^{-(2/n_2)\theta_g} \tag{44}$$

A similar proof for parallel polarization in the limit $\theta_g \to 0$ gives:

$$\Gamma_{\parallel} \approx -e^{-\left(2n_2^2/\sqrt{n_2^2-1}\right)\theta_g} \tag{45}$$

This is now generalized to account for surface roughness. Reflection coefficient for grazing angle $\theta_g$ in the specular direction is reduced by scatter from surface roughness by a factor which is an integral over the surface spatial roughness spectrum $G(\chi_x, \chi_y)$ ( 9.6.3) in [32]):

$$|V_c| \approx 1 - 2k^2 \sin\theta_g \times \iint d\chi_x d\chi_y G(\chi_x,\chi_y)\left[\sin^2\theta_g + 2\left(\frac{\chi}{k}\right)\cos\theta_g - \left(\frac{\chi}{k}\right)\right]^{1/2} \tag{46}$$

where $\chi = \sqrt{\chi_x^2 + \chi_y^2}$ in terms of spatial frequencies $\chi_x$ and $\chi_y$. We idealize the canyon wall as a corrugated surface, having roughness along the horizontal dimension, representing vertical door jams in interior corridors and exterior pillars, vertical window well edges in exterior walls. For 1-D rough surface, $G(\chi_x,\chi_y) = G(\chi_x)\delta(\chi_y)$. For grazing incidence $(\theta_g \ll 1)$, large scale roughness $\chi \ll k$, (46) is approximated as:

$$|V_c| \approx 1 - 2k^2\theta_g \sqrt{\frac{2}{k}} \int_{-\infty}^{\infty} d\chi_x G(\chi_x)\sqrt{|\chi_x|} \tag{47}$$

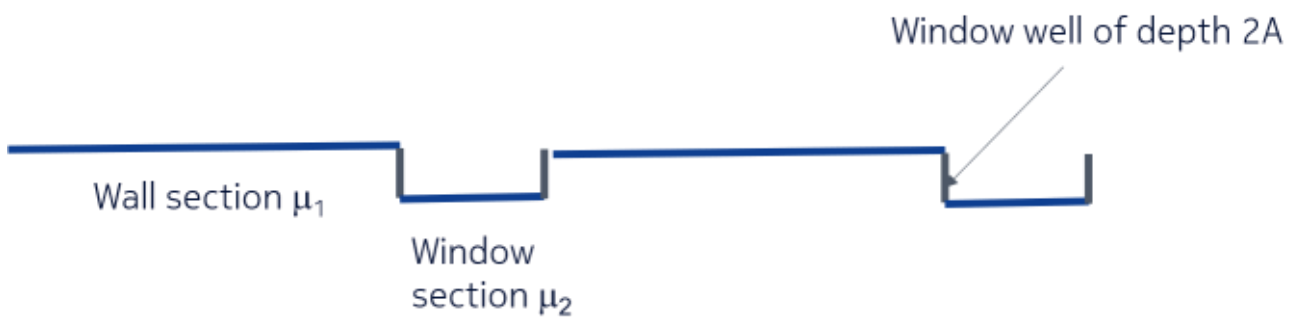

Fig. 17. Top view of a canyon boundary wall as a corrugated 2-state surface, with alternating wall sections and window well sections of average width $1/\mu_1$ and $1/\mu_2$, respectively.

The wall surface represented as a 2-state random telegraph signal, Fig. 17, (wall-window well for exterior walls and wall/doorway in interior corridors) has spatial correlation [23]

$$R(x) = \eta^2 + 4A^2 p_1 p_2 e^{-(\mu_1-\mu_2)|x|} \tag{48}$$

parametrized by amplitude $A$ (half-depth of window well), state probabilities $p_1$, $p_2$, mean $\eta = A(p_1 - p_2)$ and state transition rates $\mu_1, \mu_2$ (inversely related to average window width and window separation, respectively). The spectrum $G(\chi_x)$ is the Fourier transform of (48) with respect to *x*:

$$G(\chi_x) = \eta^2 \delta(\chi_x) + 4A^2 p_1 p_2 \frac{1}{2\pi} \frac{2|\mu_1+\mu_2|}{|\mu_1+\mu_2|^2 + \chi_x^2} \tag{49}$$

Substituting (49) into (47) and evaluating the integral leads to

$$|V_c| \approx 1 - 16k^{3/2} A^2 p_1 p_2 \sqrt{\mu_1+\mu_2}\,\theta_g \approx e^{-16k^{3/2}A^2 p_1 p_2 \sqrt{\mu_1+\mu_2}\,\theta_g} \tag{50}$$

Total reflection coefficient is taken as a combination of reflection loss from a smooth dielectric (44) and scatter from roughness (50):

$$|\Gamma| = |\Gamma_0||V_c| \approx e^{-(2/n_2)\theta_g} e^{-16k^{3/2}A^2 p_1 p_2 \sqrt{\mu_1+\mu_2}\,\theta_g} \tag{51}$$

## Appendix II: Path Gain from Free Space into Diffuse Halfspace

The canonical problem considered here is the propagation between a base station placed in free space and a terminal immersed in a diffusely scattering medium, with the two regions separated by a planar boundary with a (field) transmission coefficient *T*, illustrated in Fig. 18.

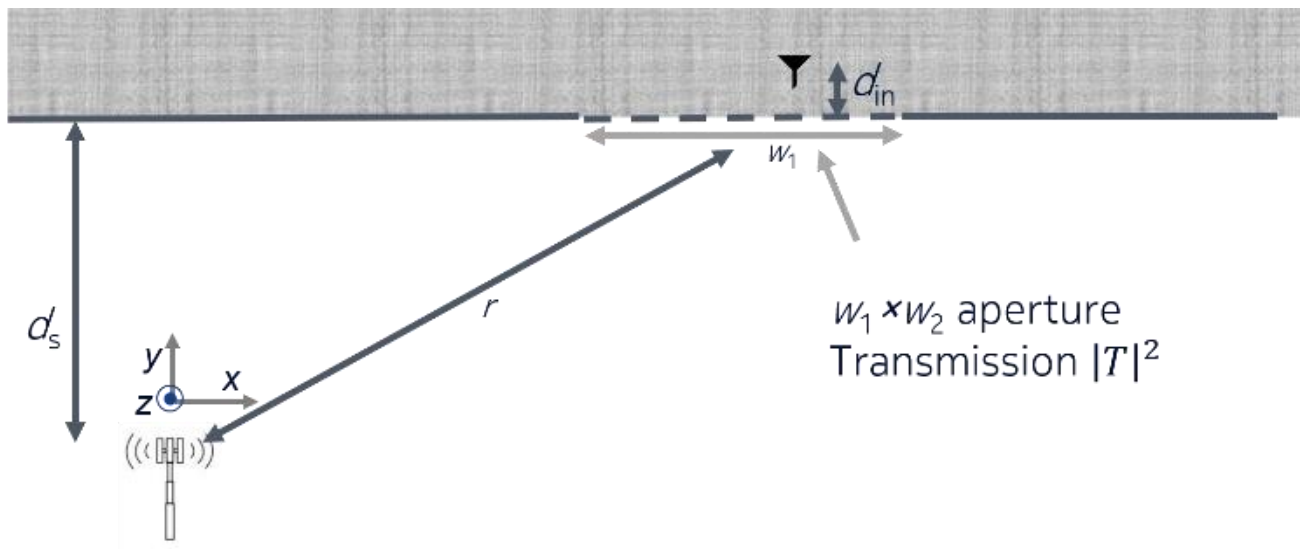

Fig. 18. Diffuse halfspace (top) illuminated by a source from free space.

The field $G_{OI}(\mathbf{r},\mathbf{r}_s)$ at $\mathbf{r}_s$ in free half-space, is related [19] to the field $G_{in}(\mathbf{r},\mathbf{r}')$ and its normal derivative $\partial G_{in}(\mathbf{r},\mathbf{r}')/\partial n'$ acting as secondary sources at intermediate locations $\mathbf{r}'$ on the boundary between the two regions:

$$G_{OI}(\mathbf{r},\mathbf{r}_s) = \iint dA \left( G_{in}(\mathbf{r},\mathbf{r}') \frac{\partial G(\mathbf{r}',\mathbf{r}_s)}{\partial n'} - G(\mathbf{r}',\mathbf{r}_s) \frac{\partial G_{in}(\mathbf{r},\mathbf{r}')}{\partial n'} \right). \tag{52}$$

In (52), $dA$ is the differential surface area, field $G_{in}(\mathbf{r},\mathbf{r}')$ is the field at the boundary point $\mathbf{r}'$ due to source at $\mathbf{r}$ in diffuse region and $G(\mathbf{r}',\mathbf{r}_s) = G_{LOS}(\mathbf{r}',\mathbf{r}_s)$ is a field in free space due to a point source at $\mathbf{r}_s$, satisfying the Helmholtz equation (2) and the boundary condition at the interface between free space and diffuse scattering half-space:

$$G_{LOS}(\mathbf{r}',\mathbf{r}_s) = \frac{e^{ik|\mathbf{r}'-\mathbf{r}_s|}}{4\pi|\mathbf{r}'-\mathbf{r}_s|} + \Gamma \frac{e^{ik|\mathbf{r}'-\mathbf{r}_i|}}{4\pi|\mathbf{r}'-\mathbf{r}_i|}, \tag{53}$$

consisting of a spherical wave incident on the surface and a reflected wave, appearing to emanate from the image source at $\mathbf{r}_i$. The boundary reflection coefficient $\Gamma \approx -1$ for low grazing angles, approximately valid for grazing incidence over a broad range of dielectric materials and for both polarizations. This leads to $G_{LOS}(\mathbf{r}',\mathbf{r}_s) = 0$ on the boundary, eliminating the 2nd term in the integrand of (52).

The problem is thus transformed into a problem where there is a distributed equivalent source, namely a "hot" wall, illuminated by a field launched from a diffuse interior and radiating into free space:

$$G_{OI}(\mathbf{r},\mathbf{r}_s) = \iint dA G_{in}(\mathbf{r}',\mathbf{r}) \left.\frac{\partial G_{LOS}(\mathbf{r}',\mathbf{r}_s)}{\partial y'}\right|_{y'=y_w} \tag{54}$$

With normal gradient of the free space $G_{LOS}(\mathbf{r}',\mathbf{r}_s)$

$$\left.\frac{\partial G_{LOS}}{\partial y'}\right|_{y'=0} = \frac{2ikd_s}{|r_s - r'|} \frac{e^{ik|r_s-r'|}}{4\pi|r_s - r'|}. \tag{55}$$

The stochastic field $G_{in}(\mathbf{r}',\mathbf{r})$ at location $\mathbf{r}'$ on the wall due to a terminal at depth $d_{in}$ at indoor location $\mathbf{r}$ has undergone scattering and absorption. It is expressed here in terms of intensity $F(\mathbf{r}',\mathbf{r})$ and a random small-scale ``Rayleigh texture" $\xi(\mathbf{r}',\mathbf{r})$ [21][30], using [34][35]:

$$\begin{aligned} &G_{in}(\mathbf{r}',\mathbf{r}) = F(\mathbf{r}',\mathbf{r})^{1/2} \xi(\mathbf{r}',\mathbf{r}) \\ &\xi(\mathbf{r}',\mathbf{r}) \sim CN(0,1), \\ &\left\langle \xi(\mathbf{r}',\mathbf{r}) \xi^*(\mathbf{r}'',\mathbf{r}) \right\rangle = \frac{2J_1(k|\mathbf{r}'-\mathbf{r}''|)}{k|\mathbf{r}'-\mathbf{r}''|} \approx \frac{4\pi}{k^2} \delta(\mathbf{r}'-\mathbf{r}''). \end{aligned} \tag{56}$$

The field $G_{in}(\mathbf{r}',\mathbf{r})$ is thus a zero-mean complex Gaussian process, spatially white in the integration surface (as indicated by the autocorrelation $\delta(\mathbf{r}'-\mathbf{r}'')$), approximating a surface field of effective coherence area $4\pi/k^2 = \lambda^2/\pi$ [35].

The field intensity $F(\mathbf{r}',\mathbf{r}_\mathrm{r}) = \left\langle \left| G_\mathrm{in}(\mathbf{r}',\mathbf{r}) \right|^2 \right\rangle$ is obtained from the diffusion solution [34] to the radial power flux $\frac{\partial}{\partial r_\mathrm{in}}\left(-\frac{e^{-\kappa r_\mathrm{in}}}{4\pi r_\mathrm{in}}\right)(\mathrm{W/m^2})$, where $r_\mathrm{in} \equiv |\mathbf{r}'-\mathbf{r}|$, projected through angle $\theta$ (Fig. 18) onto the normal to the building wall:

$$F(\mathbf{r}',\mathbf{r}) = \frac{|T|^2}{4\pi}\frac{\partial}{\partial r_\mathrm{in}}\left(-\frac{e^{-\kappa r_\mathrm{in}}}{4\pi r_\mathrm{in}}\right)\cos\theta \qquad r_\mathrm{in} \equiv |\mathbf{r}'-\mathbf{r}| \tag{57}$$

where $\cos\theta = d_\mathrm{in}/r_\mathrm{in}$ and the additional factor $1/4\pi$ is introduced for consistency between field intensity $\left\langle \left| G(\mathbf{r}',\mathbf{r}) \right|^2 \right\rangle$ and power flux. Prior to reaching the free space half-space, the flux has undergone absorption $e^{-\kappa r_\mathrm{in}}$, spreading and wall penetration losses $|T|^2$. For radiated power of unity and in the absence of absorption, $\kappa = 0$, the radial power flux incident on the wall at $\theta = 0$ reduces to $1/4\pi r_\mathrm{in}^2$, as expected from conservation of power.

Substituting (55) and (56) into (54), allows evaluation of $\left\langle \left| G_\mathrm{OI}(\mathbf{r},\mathbf{r}_s) \right|^2 \right\rangle$ as

$$\left\langle \left| G_\mathrm{OI}(\mathbf{r},\mathbf{r}_s) \right|^2 \right\rangle = 4\iint dA\, F(r)\frac{4\pi}{k^2}\frac{k^2 d_\mathrm{s}^2}{(4\pi)^2 |r_s - r'|^4} \tag{58}$$

Situation of most common interest is when the distance from the center of the "hot" wall region $\mathbf{r}_0'$ to the source in free space is much greater than to the terminal in diffuse interior region, $r \equiv |\mathbf{r}_\mathrm{s} - \mathbf{r}_0'| \gg r_\mathrm{in}$. In that case, substituting (57) into (58) leads to:

$$\left\langle \left| G_\mathrm{OI}(\mathbf{r},\mathbf{r}_s) \right|^2 \right\rangle = \frac{4 d_s^2 |T|^2}{4\pi r^4}\iint dA' \frac{\partial}{\partial r_\mathrm{in}}\left(-\frac{e^{-\kappa r_\mathrm{in}}}{(4\pi)^2 r_\mathrm{in}}\right)\frac{d_\mathrm{in}}{r_\mathrm{in}} \tag{59}$$

when the size of the "hot" wall is much greater than *r*, integration region in (59) may be first taken as unlimited, convenient to express in polar coordinates:

$$\begin{aligned}\left\langle \left| G_\mathrm{OI}(\mathbf{r},\mathbf{r}_s) \right|^2 \right\rangle &= \frac{4 d_\mathrm{s}^2 |T|^2}{4\pi r^4}\int_0^{2\pi} d\phi' \int_0^\infty \rho' d\rho' \frac{d_\mathrm{in}}{r'}\frac{\partial}{\partial r'}\left(-\frac{e^{-\kappa r'}}{(4\pi)^2 r'}\right) \\ &= \frac{4 d_\mathrm{s}^2 |T|^2}{4\pi r^4} 2\pi \int_{d_\mathrm{in}}^\infty r' dr' \frac{d}{r'}\frac{\partial}{\partial r'}\left(-\frac{e^{-\kappa r'}}{(4\pi)^2 r'}\right)\end{aligned} \tag{60}$$

using $r' = \sqrt{d_\mathrm{in}^2 + \rho'^2}$, $r'dr' = \rho' d\rho'$. Evaluating (60), and substituting into (9), produces a remarkably simple form for average received power

$$P = \frac{\lambda^2 d_\mathrm{s}^2 |T|^2 e^{-\kappa d_\mathrm{in}}}{8\pi^2 r^4} \tag{61}.$$

When the effective secondary source field region at the boundary facing the street/corridor is limited to a $w_1 \times w_2$ rectangular opening through a high loss wall (e.g. a window/door in a concrete/plywood wall, especially in mm wave bands), the area integral in (59) may be evaluated in rectangular coordinates, using $r' = \sqrt{d_\mathrm{in}^2 + x'^2 + y'^2}$ and $e^{-\kappa r'} \approx e^{-\kappa d_\mathrm{in}}$ for $\kappa \ll 1/d_\mathrm{in}$ :

$$\begin{aligned} P &\approx \frac{\lambda^2 d_\mathrm{s}^2 |T|^2 e^{-\kappa d_\mathrm{in}}}{16\pi^3 r^4}\int_{-w_1/2}^{w_1/2} dx' \int_{-w_2/2}^{w_2/2} dy' \frac{d_\mathrm{in}}{\left(d_\mathrm{in}^2 + x'^2 + y'^2\right)^{3/2}} \\ &= \frac{\lambda^2 d_\mathrm{s}^2 e^{-\kappa d_\mathrm{in}} |T|^2 \tan^{-1}\left(\frac{w_1 w_2}{2 d_\mathrm{in}\sqrt{4 d_\mathrm{in}^2 + w_1^2 + w_2^2}}\right)}{4\pi^3 r^4}\end{aligned} \tag{62}$$

Equation (62) may be rewritten as

$$P \approx \frac{\lambda^2 d_\mathrm{s}^2 e^{-\kappa d_\mathrm{in}} T_\mathrm{eff}}{8\pi^2 r^4} \tag{63}$$

In terms of the effective boundary transmission coefficient, defined for special cases of the aperture

$$\begin{aligned} T_\mathrm{eff} &= |T|^2 \frac{2}{\pi}\tan^{-1}\left(\frac{w_1 w_2}{2 d_\mathrm{in}\sqrt{4 d_\mathrm{in}^2 + w_1^2 + w_2^2}}\right), \text{ aperture } w_1 \times w_2 \\ &= |T|^2 \frac{2}{\pi}\tan^{-1}\left(\frac{w_1}{2 d_\mathrm{in}}\right), \quad \text{"street" of width } w_1 \\ &= |T|^2, \quad \text{unbounded aperture}\end{aligned} \tag{64}$$

The 2nd and 3rd line expressions in (64) are obtained as limiting cases of the 1st line expression by setting $w_2 \to \infty$ and $w_1, w_2 \to \infty$, respectively. Substituting $T_\mathrm{eff} = |T|^2$ from the last line in (64) into (63) reproduces (61), as it should for an unbounded aperture.

$|T|^2$ is the power transmission coefficient for the material covering the effective aperture. The material might be glass for a window, wood for a closed door or air ($|T|^2 = 1$) for a street viewed from above. For a more complex boundary with multiple apertures, e.g. a terminal antenna deep inside a concrete building with windows, the effective transmission coefficient may be modeled as a mixture of wall and window transmission coefficients weighted by corresponding fraction of the overall building façade [23],[3]:

$$T_\mathrm{eff} = p_\mathrm{window}\left|T_\mathrm{window}\right|^2 + \left(1 - p_\mathrm{window}\right)\left|T_\mathrm{wall}\right|^2 \tag{65}$$

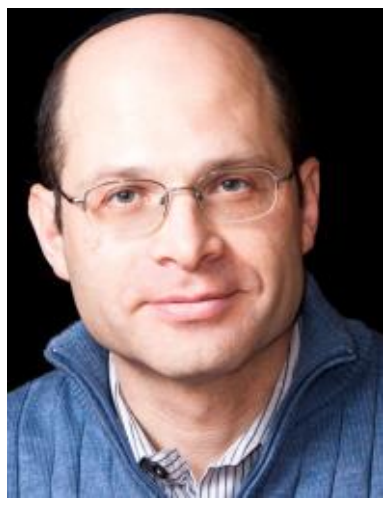

**Dmitry Chizhik** (F'14) received the Ph.D. degree in electrophysics from Polytechnic University (now NYU), Brooklyn, NY, USA. His thesis work has been in ultrasonics and non-destructive evaluation. He joined the Naval Undersea Warfare Center, New London, CT, USA, where he did research in scattering from ocean floor, geoacoustic modeling of porous media and shallow water acoustic propagation. In 1996, he joined Bell Laboratories, working on radio propagation modeling and measurements, using deterministic and statistical techniques. He has worked on measurement, modeling and channel estimation of MIMO channels. The results are used both for determination of channel-imposed bounds on channel capacity, system performance and for optimal antenna array design. His recent work has included system and link simulations of satellite and femto cell radio communications and mm wave propagation that included all aspects of the physical layer. His research interests are in acoustic and electromagnetic wave propagation and scattering, signal processing, communications, radar, sonar, and medical imaging. He is a recipient of the Bell Labs President's Award and a Distinguished Member of Technical Staff.

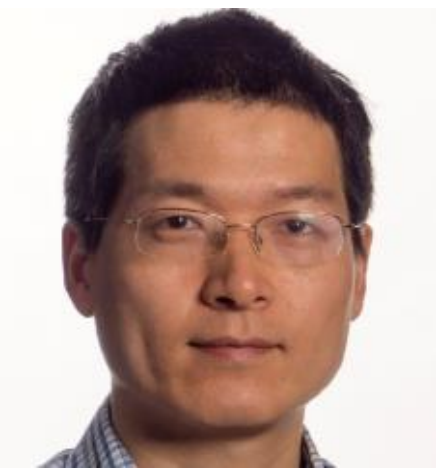

**Jinfeng Du** (Member, IEEE) received the B.Eng. degree in electronic information engineering from the University of Science and Technology of China (USTC), Hefei, China, in 2004, and the M.Sc., Tekn. Lic., and Ph.D. degrees from the Royal Institute of Technology (KTH), Stockholm, Sweden, in 2006, 2008, and 2012, respectively.

He was a Post-Doctoral Researcher with the Massachusetts Institute of Technology (MIT), Cambridge, MA, USA, from 2013 to 2015, and then joined the Bell Labs at Crawford Hill, Holmdel, NJ, USA. His research interests are in the general area of wireless communications, especially in communication theory, information theory, wireless networks, millimeter-wave propagation, and channel modeling.

Dr. Du received the Best Paper Award from IC-WCSP in October 2010, and his paper was elected as one of the Best 50 Papers in IEEE GLOBECOM 2014. He received the prestigious Hans Werthen Grant from the Royal Swedish Academy of Engineering Science (IVA) in 2011, the Chinese Government Award for Outstanding Self-Financed Students Abroad in 2012, the International PostDoc Grant from the Swedish Research Council in 2013, and three grants from the Ericsson Research Foundation.

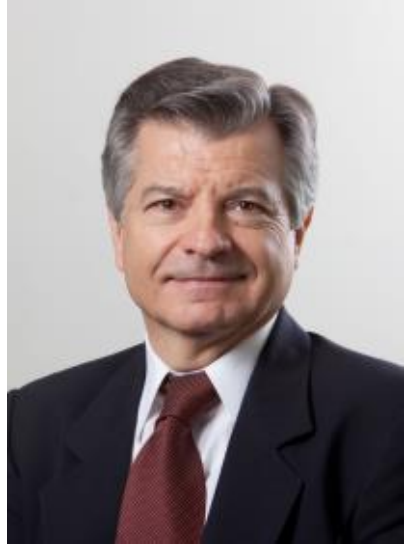

**Reinaldo A. Valenzuela** (Fellow, IEEE) received the B.Sc. degree from the University of Chile, Santiago, Chile, and the Ph.D. degree from the Imperial College London, London, U.K. He is currently the Director of the Communication Theory Department and a Distinguished Member of Technical Staff with Nokia Bell Laboratories, Murray Hill, NJ. He has published 190 articles and 44 patents. He has over 32, 600 Google Scholar citations. He is currently engaged in propagation measurements and models, MIMO/space time systems achieving high capacities using transmit and receive antenna arrays, HetNets, small cells, and next-generation air interface techniques and architectures. Dr. Valenzuela is a member of the National Academy of Engineering, a Bell Labs Fellow, a WWRF Fellow, and a Fulbright Senior Specialist. He was a recipient of the IEEE Eric E. Sumner Award, 2014 IEEE CTTC Technical Achievement Award, and 2015 IEEE VTS Avant Garde Award. He is a “Highly Cited Author” in Thomson ISI.